\documentclass[12pt,preprint]{aastex}
\newcommand{\etal}{et\,al.}
\newcommand{\lsim}{\raise0.3ex\hbox{$<$}\kern-0.75em{\lower0.65ex\hbox{$\sim$}}}
\newcommand{\gsim}{\raise0.3ex\hbox{$>$}\kern-0.75em{\lower0.65ex\hbox{$\sim$}}}
\begin{document}

\slugcomment{Astrophysical Journal, in press}

\title{Spatially Resolved PAH Emission Features in Nearby, Low Metallicity, 
Star-Forming Galaxies}




\author{Korey Haynes}
\affil{Department of Physics \& Astronomy, Macalester College, 1600 Grand 
Avenue, Saint Paul, MN 55105}
\affil{Department of Physics \& Astronomy, George Mason University,
Fairfax, VA 22030}
\email{khaynes5@gmu.edu}

\author{John M. Cannon}
\affil{Department of Physics \& Astronomy, Macalester College, 1600 Grand 
Avenue, Saint Paul, MN 55105}
\email{jcannon@macalester.edu}

\author{Evan D. Skillman} 
\affil{Department of Astronomy, University of Minnesota, Minneapolis, MN 55455}
\email{skillman@astro.umn.edu}

\author{Dale C. Jackson} 
\affil{Sandia National Laboratories, Albuquerque, New Mexico}
\email{dcjacks@sandia.gov}

\author{Robert D. Gehrz} 
\affil{Department of Astronomy, University of Minnesota, Minneapolis, MN 55455}
\email{gehrz@astro.umn.edu}

\begin{abstract}

Low-resolution, mid-infrared {\it Spitzer}/IRS spectral maps are
presented for three nearby, low-metallicity dwarf galaxies (NGC\,55,
NGC\,3109 and IC\,5152) for the purpose of examining the spatial
distribution and variation of polycyclic aromatic hydrocarbon (PAH)
emission.  The sample straddles a metallicity of 12 + log(O/H)
$\approx$ 8, a transition point below which PAH intensity empirically
drops and the character of the interstellar medium changes.  We derive
quantitative radiances of PAH features and atomic lines on both global
and spatially-resolved scales. The {\it Spitzer} spectra, combined
with extensive ancillary data from the UV through the mid-infrared,
allow us to examine changes in the physical environments and in PAH
feature radiances down to a physical scale of $\sim$50 pc. We discuss
correlations between various PAH emission feature and atomic line
radiances. The (6.2 $\mu$m)/(11.3 $\mu$m), (7.7 $\mu$m)/(11.3 $\mu$m),
(8.6 $\mu$m)/(11.3 $\mu$m), (7.7 $\mu$m)/(6.2 $\mu$m), and (8.6 $\mu$m)/(6.2 $\mu$m) PAH
radiance ratios are found to be independent of position across all
three galaxies, although the ratios do vary from galaxy to galaxy.  As
seen in other galaxies, we find no variation in the grain size
distribution as a function of local radiation field strength.
Absolute PAH feature intensities as measured by a ratio of
PAH/(24 $\mu$m) radiances are seen to vary both positionally within a
given galaxy, and from one galaxy to another when integrated over the
full observed extent of each system. We examine direct comparisons of
CC mode PAH ratios (7.7 $\mu$m)/(6.2 $\mu$m) and (8.6 $\mu$m)/(6.2 $\mu$m) to the
mixed (CC/CH) mode PAH ratio (7.7 $\mu$m)/(11.3 $\mu$m). We find little
variation in either mode, and no difference in trends between
modes. While the local conditions change markedly over the observed
regions of these galaxies, the properties of PAH emission show a
remarkable degree of uniformity.

\end{abstract}						


\section{Introduction} \label{S1}

\subsection{On The Origin of PAH Emission Features} \label{S1.1}

The mid-infrared (MIR; $\sim$3--20 $\mu$m) is a very rich spectral
regime. The Rayleigh Jean's tail of the stellar spectral energy
distribution (SED) contributes at short wavelengths, and the rise of
the warm dust continuum contributes at long wavelengths. Between them
is the large silicate extinction feature centered around 9.7
$\mu$m. Superposed on this complex continuum are numerous broad
emission features (notably at 3.3 $\mu$m, 6.2 $\mu$m, 7.7 $\mu$m, 8.6
$\mu$m, 11.3 $\mu$m and 12.6 $\mu$m) and atomic lines (e.g., [S IV] at
10.5 $\mu$m, [Ne~II] at 12.8 $\mu$m, [Ne~III] at 15.56 $\mu$m, and [S
III] at 18.71 $\mu$m, among others). The broad emission features are
attributed to vibrational transitions from polycyclic aromatic
hydrocarbon (PAH) molecules.

PAHs are planar molecules containing tens to thousands
of carbon atoms.  While the assignment of these features to specific
molecules is still an area of very active theoretical and laboratory
investigation, it is generally agreed that this type of molecule
(carbon atoms arranged in a system of fused hexagonal rings terminated
by carbon-hydrogen single bonds) produces the broad emission features
in the MIR spectral range.  The four most luminous features in
astrophysical sources appear at wavelengths of 6.2 $\mu$m, 7.7 $\mu$m,
8.6 $\mu$m and 11.3 $\mu$m \citep{Leger84}. These emission features
are produced following the absorption of energetic (typically
far-ultraviolet) photons by PAH molecules; the molecules bend and
stretch in fundamental modes that produce infrared photons as this
energy is dissipated.

There are two principal types of vibrational modes in PAH
molecules. The first involves the stretching of carbon-carbon (CC)
bonds and contributes to the features at 6.2 $\mu$m, 7.7 $\mu$m and
8.6 $\mu$m.  The second involves the bending of carbon-hydrogen (CH)
bonds; these modes can occur either in the plane (``wagging'' modes)
or out of the plane of the molecule.  The wagging mode contributes to
the 7.7 $\mu$m and 8.6 $\mu$m features, while the out-of-plane bending
mode is less energetic and contributes to the 11.3 $\mu$m feature.
For detailed discussions of these mode assignments (some of which are
still under investigation) we refer the reader to
\citet{Allamandola89}, \citet{Draine07}, \citet{Tielens08},
Bauschlicher \etal\ ({2008}\nocite{Bauschlicher08},
{2009}\nocite{Bauschlicher09}) and the many references therein.

The relative variations in intensity of these features is sensitive to
both the physical size and to the ionization state (i.e., to the
hardness of the radiation field) of the molecules.  It is generally
accepted that the smallest PAH molecules produce the
shortest-wavelength features, and that increasing the number of carbon
atoms leads to increased intensities of the longer wavelength
features.  However, this simple interpretation is clouded somewhat by
the prediction that the 7.7 $\mu$m band is produced by a mixture of
both small and large molecules \citep{Bauschlicher09}.  Evidence is
also mounting that the 6.2 $\mu$m feature may arise from PAHs with
different compositions than the longer-wavelength features, either
from the inclusion of N atoms (Bauschlicher
\etal\ {2008}\nocite{Bauschlicher08}, {2009}\nocite{Bauschlicher09})
or from the carriers being among the simplest C-based ring structures
\citep[e.g., protonated naphthalene; ][]{Ricks09}.  In this simple
model, the ratios of radiances of long-wavelength PAH features to
shorter-wavelength features provides a diagnostic of the grain size
distribution; examining the changes in the grain size distribution as
a function of local conditions is one of the primary goals of this
study.

The hardness of the radiation field also contributes to the variations
in intensity of the various PAH bands. Laboratory predictions
\citep[e.g.,][]{Allamandola99,Kim02,Bauschlicher02} suggest that
ionized PAH molecules produce stronger CC mode vibrations
\citep[e.g.,][]{Bakes01}. \citet{Bauschlicher08} finds that the
intensities of the CC modes arising from cationic PAHs are increased
by an order of magnitude or more compared to neutral species.
Observationally this manifests itself as a strengthening of the CC
modes (i.e., the 6.2 $\mu$m, 7.7 $\mu$m and 8.6 $\mu$m PAH features)
relative to the CH modes (primarily the 11.3 $\mu$m feature, although
the 8.6 $\mu$m feature is also affected to a lesser extent) in a given
source.  Quantitative studies have been undertaken to attempt to
connect these variations of PAH intensities to the physical conditions
of the gas in a variety of astrophysical environments (see further
discussion below).

\subsection{Studying PAH Emission in Nearby Galaxies} \label{S1.2}

Galaxies radiate a large percentage of their total luminosity in the
infrared, from tens of percent in star forming galaxies up to 99\% in
ultra luminous infrared galaxies \citep{Sanders88,Sanders96}. An
especially rich segment of the IR region falls between $\sim$3 $\mu$m
- 20 $\mu$m, where emission features from PAHs arise in most
star-forming galaxies.  The emission from PAHs varies from a few
percent up to 20\% of the total MIR luminosity in normal galaxies
\citep{Helou00}, but this percentage drops sharply with decreasing
metallicity \citep{Engelbracht05,Jackson06,Smith07a}. Isolating the
processes that affect PAH emission intensity allows investigation of
the physical conditions present in star forming regions
\citep{Kennicutt03}. This in turn leads to enhanced understanding of
the interplay between PAH emission and star formation, the role of
radiation field characteristics in governing PAH intensities, and the
stellar-interstellar medium connection.

Many investigations have been conducted on PAH emission in a wide
range of galaxy types, but the observations of low-metallicity
galaxies in particular have been challenging, in large part due to the
difficulties of obtaining spectra of sufficient sensitivity and
spatial resolution from low-mass (and often low surface-brightness)
galaxies. Using Infrared Array Camera (IRAC) imaging,
\citet{Engelbracht05} discovered a threshold metallicity of 12 +
log(O/H) $\approx$ 8, below which PAH intensity drops sharply, and
these results have been confirmed by a number of other studies
\citep[e.g.,][and references
  therein]{Hogg05,Ohalloran06,Jackson06,Rosenberg06,Engelbracht08,Wu10}. This
metallicity apparently signifies a fundamental change in the nature of
the interstellar medium (ISM); the molecular component of more
metal-rich galaxies is easily studied via CO observations.  However,
in systems more metal-poor than $\sim$25\% Z$_{\odot}$, CO no longer
traces H$_{\rm 2}$ accurately and the dominant molecular component by
mass remains largely unconstrained \citep{Taylor98, Leroy05}.  In this
regard, PAH observations of metal-poor galaxies are especially
important, as they provide one of our only probes of an otherwise
elusive ISM component.

Multiple factors have been found to influence the amount of infrared
emission from PAH molecules. The metal content of the host galaxy
might be expected to play an important role, since the carrier
molecules consist in part of heavy elements.  Indeed,
\citet{Rosenberg06} find a correlation between diffuse 8 $\mu$m
luminosity and metallicity for a large sample of star-forming dwarf
galaxies. The same investigation also finds a more significant
correlation between PAH luminosity and current star formation rate
(SFR) than with metallicity.  The \citet{Gordon08} study of H~II
regions in M\,101 reaches a similar conclusion, where PAH luminosities
are more closely tied to the local ionization index than to the
metallicity. These results suggest that multiple factors contribute to
the observed PAH luminosity in a given region, including process
related to both the growth and the stimulation of the molecules.

Dwarf galaxies offer canonically ``simple'' environments in which to
explore variations in the intensities of PAH features and
emissions. Their characteristically low masses make them more
susceptible than spiral galaxies to turbulence and feedback
\citep{Tenoriotagle88}, allowing recent disturbances of the galaxy's
dust and gas to be readily seen. Typical dwarfs display solid body
rotation and lack shear, preserving structures in the ISM. Dwarf
galaxies are abundant in the local universe, and were even more so in
the past. The low metallicities typical of these systems approximate
the conditions of high redshift galaxies, revealing the physical
properties of some of the most distant galaxies by using some of the
nearest as a proxy.

Previous studies of PAH emission in dwarf galaxies have sought to
explain how these molecules form, survive, and are destroyed in
low-metallicity systems. \citet{Jackson06} observed a sample of 15
Local Group dwarfs, including the 3 galaxies in the present
study. They find that both metallicity and local radiation field
properties influence the intensity of PAH emission.  They also find
that diffuse 8 $\mu$m emission (strictly a nonlinear composite of the
6.2 $\mu$m, 7.7 $\mu$m and 8.6 $\mu$m PAH features, as well as the
underlying warm dust continuum) cannot be predicted by the mass of the
galaxy alone. Another concern was whether younger systems have time to
populate their ISM with the requisite metals to grow PAHs in the first
place, but \citet{Jackson06} rules out this hypothesis. The galaxies
in their sample (and therefore ours) have all populated their
asymptotic giant branch (AGB), and therefore have the necessary metals
to grow PAHs \citep[see, for example, the discussion in][]{Tielens08},
whether or not they display PAH emission. They also dismiss outflow
being responsible for removing PAHs from low-mass galaxies since the
HI distributions for the galaxies in their sample lack evidence of
large-scale blowout. However, the relative importance of PAH
destruction remains an open issue; SNe may be able to destroy PAHs at
a rate comparable to the production rate in low-metallicity
environments.

Environmental differences in the character of PAH emission have
already been observed within galaxies using {\it Spitzer} data
\citep[e.g.,][]{Smith07a,Dale09a}. In spiral systems with large,
well-defined structures, this is to be expected. The nuclear regions,
spiral arms, inter-arm regions, and extended disks of such galaxies
can have vastly different physical conditions, as well as different
chemical compositions and SFRs. In contrast, dwarf galaxies present
surprisingly uniform metallicity throughout their high surface
brightness components \citep[e.g.,][]{Kobulnicky96,Kobulnicky97}.
While star formation regions are often scattered throughout the disks
of dwarfs, these regions appear to have only a minimal effect on the
metallicity of the surrounding interstellar gas.

The uniform metal content within a given dwarf galaxy is particularly
relevant to the issue of stimulating PAH carriers into emission.  It
has been shown that PAH emission highlights the edges of
photodissociation regions (PDRs) in spiral galaxies
\citep{Giard94,Helou01,Helou04}. As a result, it has been proposed
that radiation inside PDRs is too intense for the survival of PAH
molecules, while at large distances from PDRs the radiation field
would be too weak to stimulate the molecules.  There is also evidence
that in regions of hard and/or strong radiation fields, the type of
PAH carrier can have a marked effect on the relative PAH intensities
\citep{Compiegne07}. If radiation field strength were the only factor
affecting the emission from PAH carriers, then the uniform metallicity
ISM of dwarfs would reveal this as a correspondence between the
locations of massive stars and the presence of PAH emission.  However,
PAH emission has been observed in the quiescent ISM of dwarfs via
imaging \citep[e.g.,][]{Cannon06b,Jackson06}, including in the
galaxies observed in this study, so the physical proximity of PAHs to
PDRs cannot act alone in governing the amount of emission from PAH
molecules in the metal-poor ISM.

The spatially resolved properties of PAH emission using spectra have
not been investigated extensively for low-metallicity systems, though
many studies of diffuse emission (arising from multiple PAH features
within a bandpass) are now in the literature (see above). In this
paper, we examine the intensities and spatial variations of several
PAH features in three low-metallicity dwarf galaxies. This approach
allows one to separate and study the individual PAH bands contained
within the comparatively broad IRAC bandpasses.  Our interest is in
both how the relative intensities of these features change from one
galaxy to the next, and more importantly, within a given galaxy. A
comparison of the ratios of PAH bands within a galaxy should be
instructive as to how environmental factors and PAH intensities are
related. As one effort to trace these environmental factors, maps of
[Ne~II] emission are also extracted from our spectral maps. These
images reveal the morphology of ionized gas and offer insights into
how local radiation field properties affect the relative fluxes of the
various PAH bands.

The three galaxies chosen for this sample are NGC\,55, NGC\,3109, and
IC\,5152 (see Table~\ref{galaxysample}). All three are low-metallicity
(12\% Z$_{\odot}$ $\le$ Z $\le$ 49\% Z$_{\odot}$; see below) dwarf
galaxies located within 2 Mpc. These systems were chosen for their
proximity (maximizing spatial resolution) and for their ongoing star
formation.  Further, the metallicities of the galaxies in this sample
bracket the aforementioned empirical PAH transition metallicity.
While our sample size is modest, the spatially resolved data for each
system provide new insights into the variations of PAH intensities in
metal-poor environments.

This research uses the Infrared Spectrograph \citep[IRS;][]{Houck04}
Short-Low module to create spectral maps of the high IR surface
brightness regions of the sample galaxies. IRS is able to cleanly
separate the 6.2 $\mu$m, 7.7 $\mu$m, and 8.6 $\mu$m PAH features, as
well as to differentiate between the PAH features and the underlying
warm dust continuum.  We examine the radiances of PAH features across
the observed regions of the target galaxies. From these measurements,
combined with extensive ancillary data from the UV through the
mid-infrared, it is possible to correlate emission from various
gaseous and stellar components.  We can then constrain some of the
factors suspected of affecting the intensity of PAH emission (e.g.,
radiation field hardness, PAH molecule size distribution).

\section{Observations and Data Reduction}
\label{S2}

Data were obtained using {\it Spitzer}/IRS for general observer (GO)
Program 40457 (PI Skillman). Spectra were acquired with the Short-Low
(SL) slit (5-15 $\mu$m) for all three targets. Mapping used parallel
steps of size 28.35\arcsec, and perpendicular steps of size
1.85\arcsec. NGC\,55 was observed for a total of 7 hours, using one
cycle of ramp duration 14 seconds on 6 parallel and 60 perpendicular
steps. 6.2 hours were required to observe IC\,5152 with one 60 second
ramp on 2 parallel steps and 50 perpendicular steps. NGC\,3109 also
required 6.2 hours for two cycles of 60 second ramps on one parallel
step and 70 perpendicular steps. Observations were completed between
November 2007 and January 2008. IRS mapping footprints can be seen in
Figure \ref{IRSoverlay}, where we overlay the IRS spectral map
placement on IRAC 8 $\mu$m images. Note that we targeted the areas of
highest IR surface brightness.

Basic calibrated data (BCD) from the {\it Spitzer} pipeline has been
flat-fielded and converted to specific intensity units (MJy
sr$^{-1}$). We use the CUbe Builder for IRS Spectral Mapping
\citep[CUBISM;][]{Smith07b} for much of the analysis in this
investigation.  First, the data were cleaned using a 4-sigma cut,
removing hot pixels.  Further cleaning was then done by hand to assure
data quality.

For IC\,5152 and NGC\,3109, dedicated sky observations were used for
background subtraction. NGC\,55 had a sufficient number of off-target
BCD sets that could be used for background subtraction. These
off-target BCDs were judged to be cleaner than the dedicated sky data,
since more and cleaner astronomical observation requests (AORs) from
the target data set were available for background subtraction than
from the dedicated sky observations.

For stitching together spectral data from the first and second order
of the slit, a low order polynomial was fitted to the SL2 data in the
region of overlap. Specific intensity values were then averaged in the
region of overlap. All data were then smoothed over three
pixels. Fitting errors were small, so total errors are presented,
except in the region of overlap, where the fitting errors were
significant. For NGC\,55, NGC\,3109, and IC\,5152, average errors were
2\%, 17\%, and 10\% respectively. The only areas of significantly
higher error were in the region of overlap, where errors ranged up to
three times higher than the averages outside of the overlap region.

The program PAHFIT \citep{Smith07b} was used to fit the spectra and to
extract numerical values describing PAH features and emission
lines. PAHFIT was trained on a number of high S/N ratio galaxies, and
uses a simple, physically motivated model that includes dust
continuum, starlight, emission lines and bands, and extinction. Most
components are fixed, including starlight temperature (5000 K) and
central wavelengths and widths of dust features. The dust continuum is
based on eight components represented by modified blackbodies at a
range of temperatures between 35 and 300 K. These components are
allowed to be zero, though not negative. The extinction is given as a
power law with silicate features peaking at 9.7 $\mu$m and 18
$\mu$m. The line feature wavelengths are allowed to vary by 0.05
$\mu$m, and line widths are allowed to vary by 10\% from their default
values. Spectral lines are represented by Gaussian profiles, while
dust features are represented by Drude profiles, both individual and
blended. Because of the highly blended nature of many of the PAH
features, the PAHFIT ``Main Power Feature'' program was used to
combine the most heavily blended features into complexes by adding
individual features across a specified wavelength interval.  PAHFIT is
designed to use combined SL and LL {\it Spitzer} data in order to
accurately reconstruct the continuum emission; unfortunately, LL data
are unavailable for our sample galaxies. Therefore, while the total
continuum appears to fit the spectra quite well, some caution should
be used in interpreting the continuum fits themselves (especially at
wavelengths outside of the 5.5--14.5 $\mu$m range covered by the IRS
SL1 and SL2 slits).

PAHFIT fits the spectra by minimizing the global $\chi$$^{2}$, and
returns statistical uncertainties. Global $\chi$$^{2}$ values for
NGC\,55, NGC\,3109, and IC\,5152 are 59.122, 5.76, and 5.20,
respectively. Average uncertainties in dust features were 2\%, 17\%,
and 7\% for the global fits for NGC\,55, NGC\,3109, and IC\,5152,
respectively.

In addition to 1D spectra, 2D spectral maps from various wavelength
regions were also generated using CUBISM's mapping feature. This
procedure allows the user to specify ``peak'' and ``continuum''
wavelength intervals. CUBISM then subtracts the specified continuum
from the peak, and creates a map of the resulting emission. For maps
covering spectral regions between 7.4 $\mu$m and 7.6 $\mu$m, the
overlap region between SL1 and SL2, a more detailed method was
needed. In this case, maps were extracted separately in each slit and
averaged together in the area of overlap before being combined.

\section{Global Comparisons}
\label{S3}
\subsection{Galaxy Properties}
\label{S3.1}

As noted above, the galaxies in our sample were selected based on
proximity and the presence of ongoing star formation.  These three
dwarf irregulars span a modest range in metallicity ($\sim$0.6 dex),
but larger ranges in SFR, dynamical mass and absolute magnitude (Table
\ref{galaxysample}). The ISM oxygen abundances have been determined
through observations of H~II regions.  The temperature sensitive
[O~III] $\lambda$4363 line was detected in each system, resulting in
reliable abundance determinations. These abundances are
12\,$+$\,log(O/H) = 8.4\,$\pm$\,0.1 for NGC~55 \citep{Webster83},
7.92\,$\pm$\,0.07 for IC~5152 \citep{Lee03a}, and 7.74\,$\pm$\,0.33
for NGC\,3109 \citep{Lee03b}.

\begin{figure}
\begin{center}
\includegraphics[width=.63\textwidth]{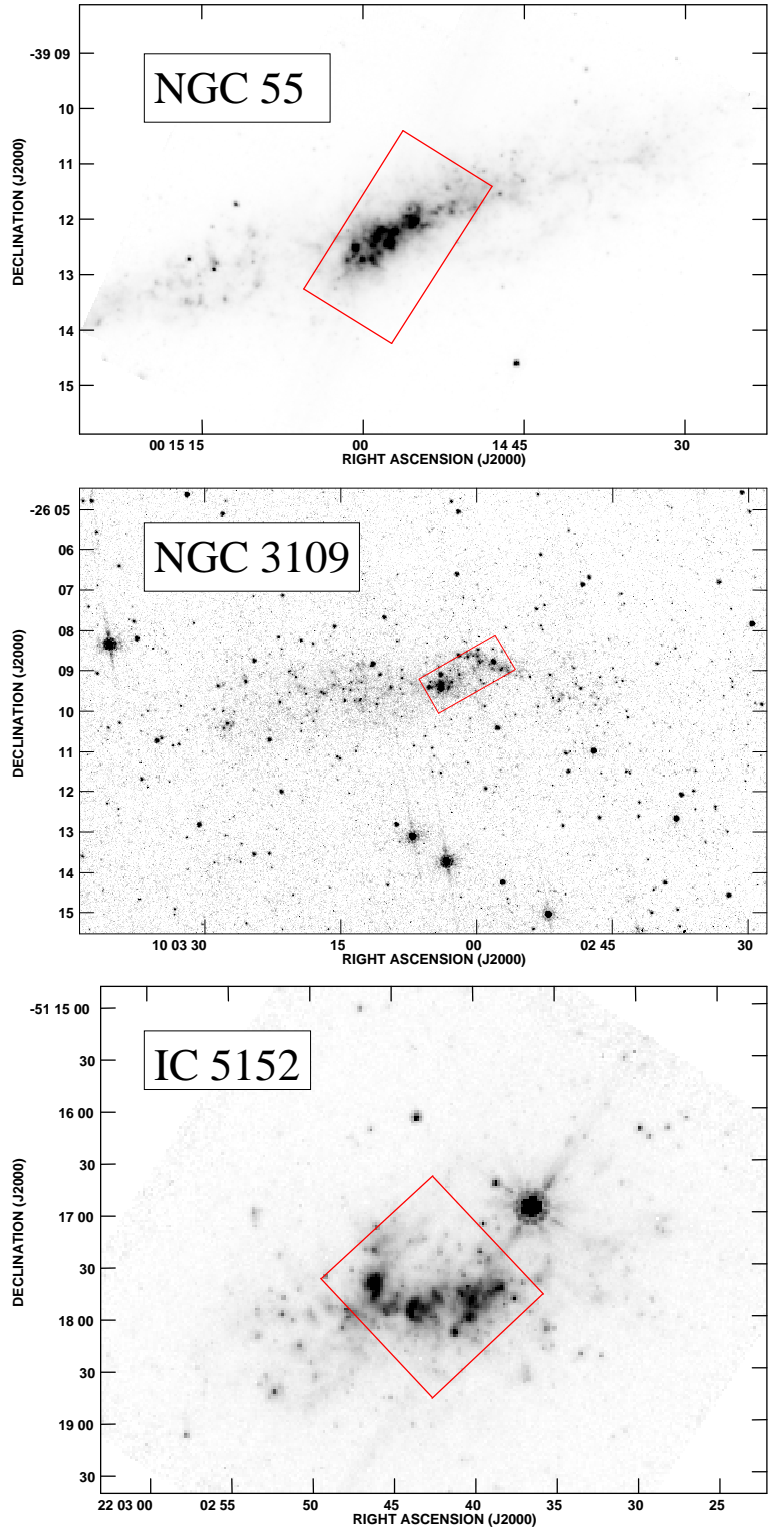}
\end{center}
\caption{IRAC 8 $\mu$m images are shown for each of the target
  galaxies with the region observed by IRS boxed in red. Images have
  arbitrary scaling to highlight features. Galactic foreground stars
  are prominent in NGC\,3109 and IC\,5152 (see in particular the
  object near IC\,5152, $\alpha \approx$ 22:02:36, $\delta \approx$
  -51:16:55).}
\label{IRSoverlay}
\end{figure}

Previous investigations have studied the luminosity of the 8 $\mu$m
PAH feature via imaging.  Typically the IRAC band 1 or 2 images (at
3.6 $\mu$m and 4.5 $\mu$m, respectively) are assumed to accurately
represent the underlying stellar continuum; these images are scaled
and subtracted from the IRAC 8 $\mu$ image (see detailed discussion in
{Jackson \etal\ 2006}\nocite{Jackson06}).  The resulting diffuse 8
$\mu$m emission (which includes contributions from the 6.2 $\mu$m, 7.7
$\mu$m and 8.6 $\mu$m PAH features) in these systems scales
approximately with the metallicity.  NGC\,3109 has the lowest
gas-phase abundance ($\sim$12\% Z$_{\odot}$) and the lowest total
diffuse 8 $\mu$m flux density (0.06 Jy).  IC\,5152 is slightly more
metal-rich ($\sim$18\% Z$_{\odot}$) and harbors a larger total 8
$\mu$m flux density (0.16 Jy).  Moving above the threshold metallicity
(12 $+$ log(O/H) $\simeq$ 8.0; {Engelbracht
  \etal\ 2005}\nocite{Engelbracht05}), NGC\,55 ($\sim$49\%
Z$_{\odot}$) is an order of magnitude brighter in the IRAC 8 $\mu$m
band (1.4 Jy).  Note by examining Figure~\ref{IRSoverlay} that our IRS
spectral maps cover only the regions of highest surface brightness 8
$\mu$m emission; the IRS field of view encompasses 41\%, 18\%, and
57\% of the intensity in the IRAC 8 $\mu$m field of view for NGC\,55,
NGC\,3109, and IC\,5152, respectively (with no foreground star
correction or continuum subtraction applied).

While the scaling of PAH emission (as measured by the diffuse 8 $\mu$m
flux density) with metallicity is expected based on recent empirical
results, the comparisons with other global parameters are more
ambiguous.  The system with the smallest dynamical mass and current
SFR (IC\,5152) is a factor of $\sim$2 more luminous in the IRAC 8
$\mu$m band than NGC\,3109.  While the current SFRs in these two
systems differ only at the $\sim$10\% level, it is perhaps more
surprising that NGC\,3109 is $\sim$16 times more massive than IC\,5152.
Moving further upward in mass, NGC\,55 is the most massive and
metal-rich system in the sample.  Its current SFR and diffuse 8 $\mu$m
flux density are both more than an order of magnitude larger than in
NGC\,3109, although the mass and metallicity are only larger by a
factor of $\sim$3.  Taken together, this discussion highlights the
complexity of these multiple inter-related properties.  While
statistical correlations exist between some parameters in large
samples \citep[e.g.,][]{Rosenberg06}, it is apparent that individual
systems are complex composites of multiple factors.  It is our aim to
investigate these variations within three selected systems on a
spatially resolved basis in the present investigation.  We thus
hereafter use the term ``global'' to identify results or properties
obtained by integrating spectra over the IRS fields of view as shown
in Figure \ref{IRSoverlay}.  These global results are separated from
those attained by individually integrating over smaller regions within
the field of view (see \S~\ref{S4}).

\subsection{Spectra}
\label{S3.2}
\subsubsection{Comparing NGC\,55, NGC\,3109 and IC\,5152}
\label{S3.2.1}

\begin{figure}
\begin{center}
\includegraphics[width=.8\textwidth,angle=90]{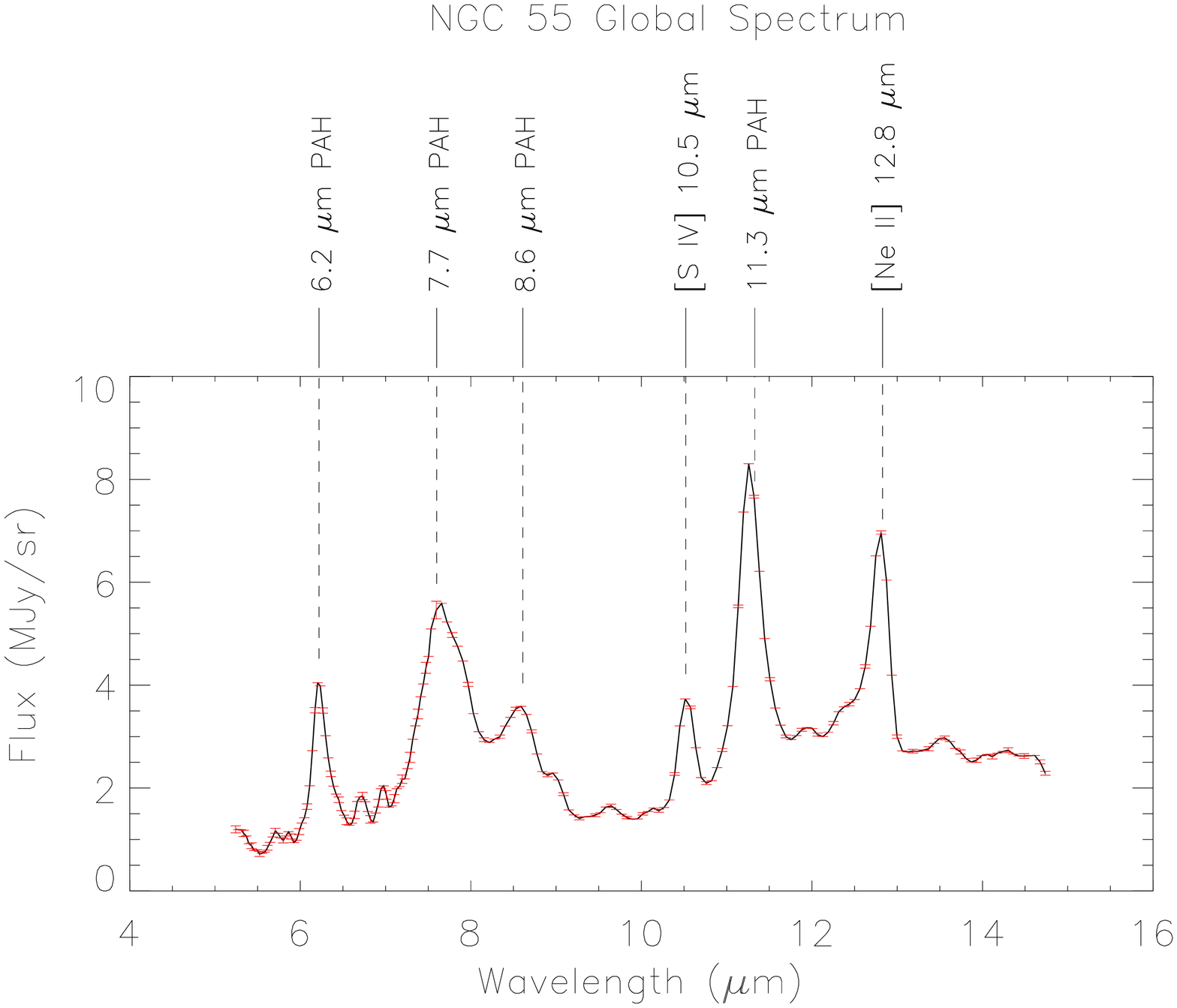}
\end{center}
\caption{Global IRS spectrum for NGC\,55, with prominent emission
  lines and PAH features labeled. Errors are shown in red, and are
  largest in the region of overlap between the SL1 and SL2 slits,
  between 7.4 $\mu$m and 7.7 $\mu$m.}
\label{idlglobalspec55}
\end{figure}

\begin{figure}
\begin{center}
\includegraphics[width=.8\textwidth,angle=90]{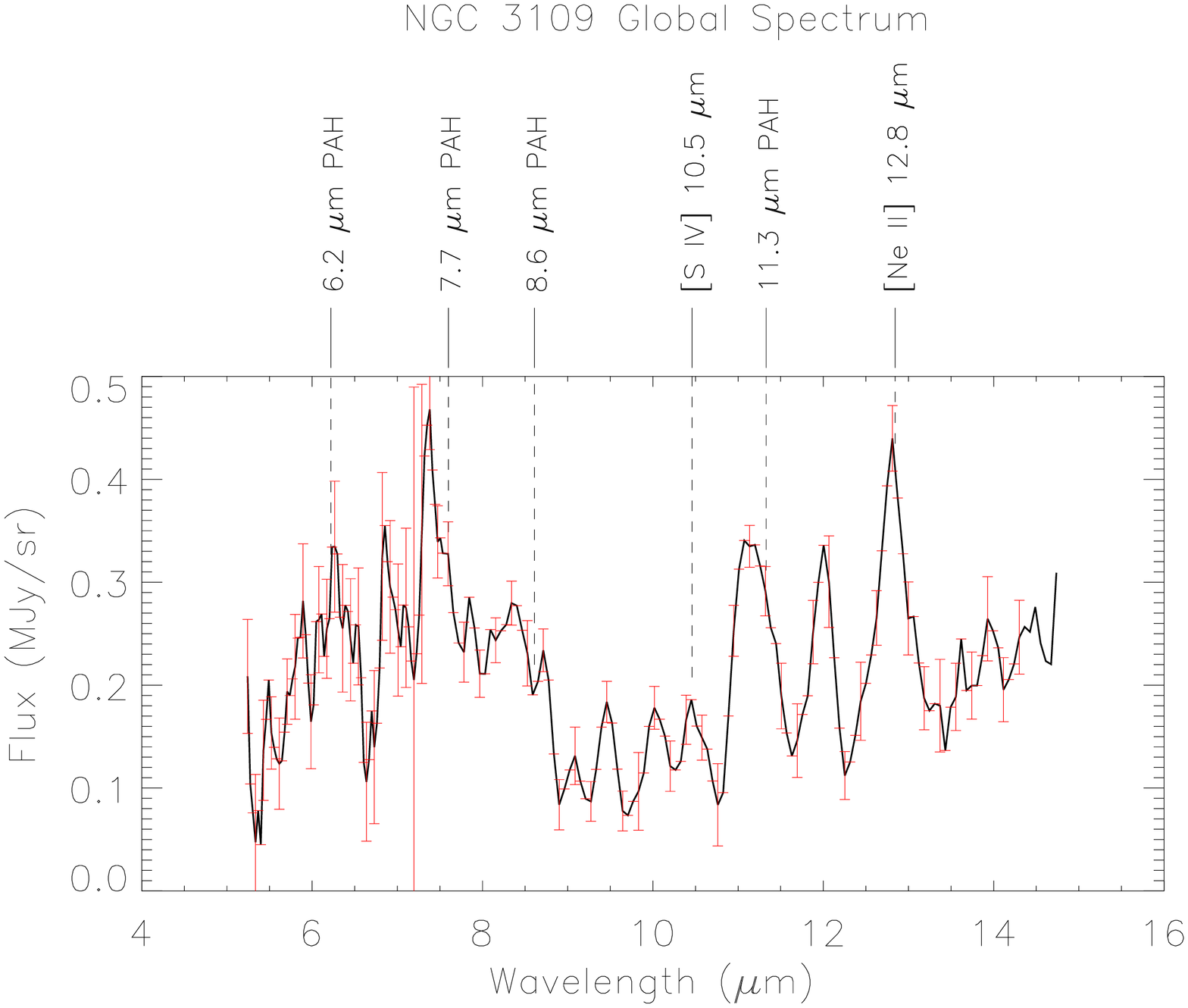}
\end{center}
\caption{Global IRS spectrum for NGC\,3109, with prominent emission
  lines and PAH features labeled. Errors are shown in red, and are
  largest in the region of overlap between the SL1 and SL2 slits,
  between 7.4 $\mu$m and 7.7 $\mu$m.}
\label{idlglobalspec3109}
\end{figure}

\begin{figure}
\begin{center}
\includegraphics[width=.8\textwidth,angle=90]{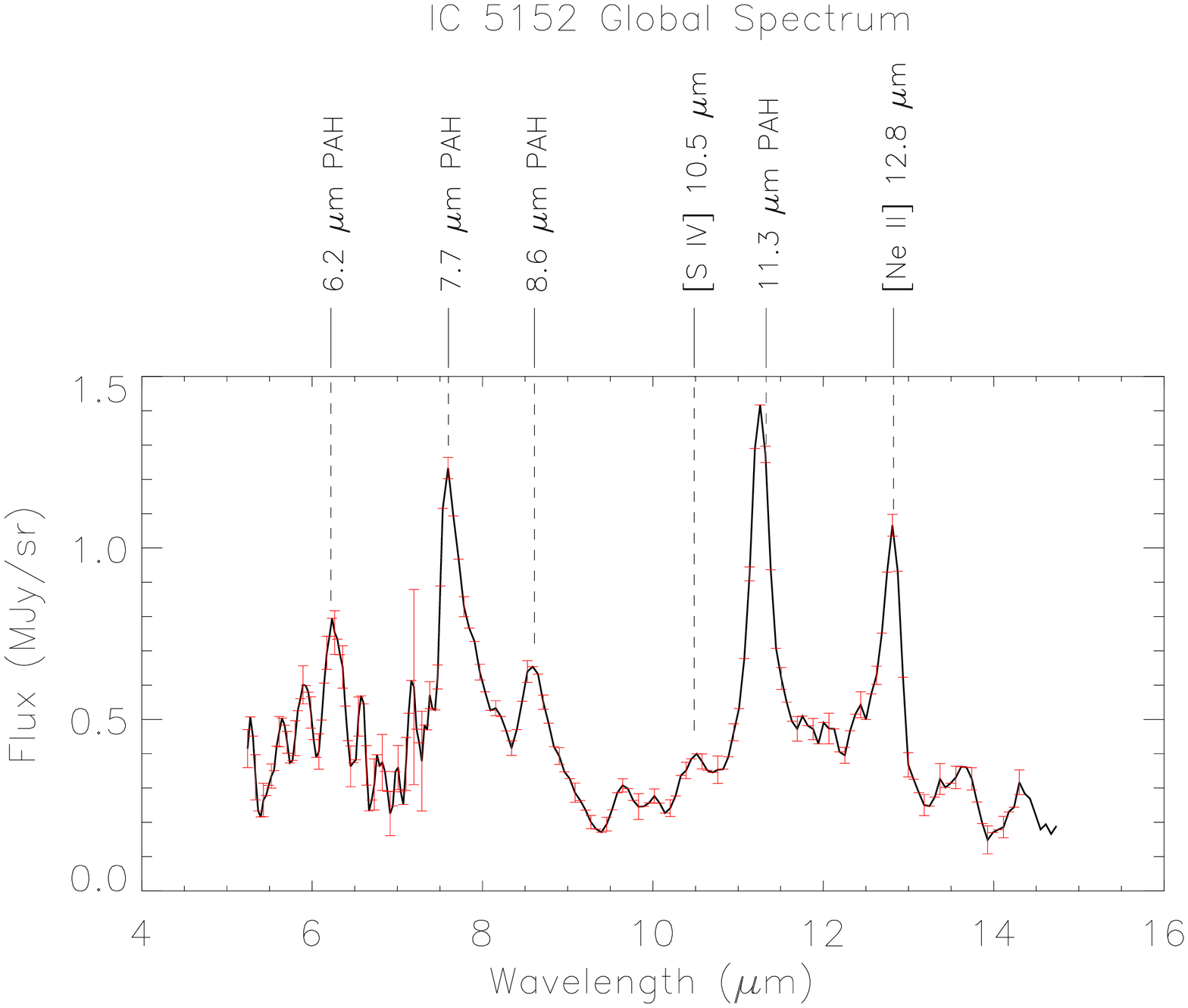}
\end{center}
\caption{Global IRS spectrum for IC\,5152, with prominent emission
  lines and PAH features labeled. Errors are shown in red, and are
  largest in the region of overlap between the SL1 and SL2 slits,
  between 7.4 $\mu$m and 7.7 $\mu$m.}
\label{idlglobalspec5152}
\end{figure}

The global spectra shown in Figures \ref{idlglobalspec55},
\ref{idlglobalspec3109}, and \ref{idlglobalspec5152} were created by
integrating the IRS specific intensity over the entire observed area
of the galaxy for NGC\,55, NGC\,3109, and IC\,5152,
respectively. While these spectra vary significantly in total
intensity and signal to noise ratio, the overall shapes show similar
emission patterns. The most prominent features in all three galaxies
are the 7.7 $\mu$m PAH feature (somewhat blended with the 8.6 $\mu$m
PAH feature), the 11.3 $\mu$m PAH feature, and the blended feature
composed of [Ne~II] emission at 12.8 $\mu$m and the 12.6 $\mu$m PAH
feature. The [S IV] feature at 10.5 $\mu$m is prominent in NGC\,55,
weak in IC\,5152, and undetectable in NGC\,3109.

\begin{figure}
\plotone{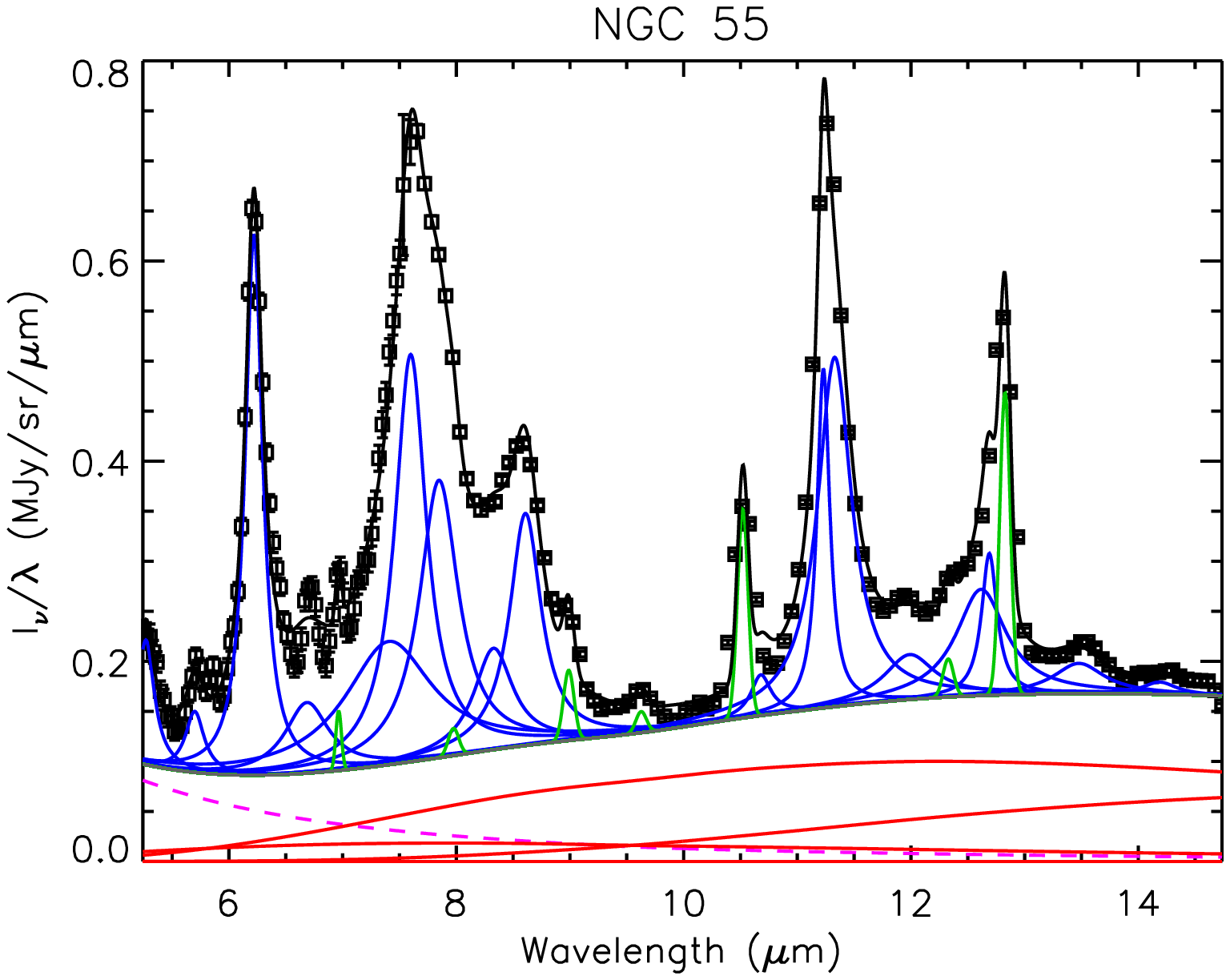}
\caption{Decomposed global IRS spectrum for NGC\,55 produced by
  PAHFIT. The red lines represent thermal dust continuum components at
  various temperatures, the dashed magenta line the stellar continuum,
  the gray line the total (dust + stellar) continuum. The blue line
  represents the PAH features, the green line the unresolved atomic
  and molecular spectral lines, and the black line the full fitted
  model.}
\label{pahfit55}
\end{figure}

\begin{figure}
\plotone{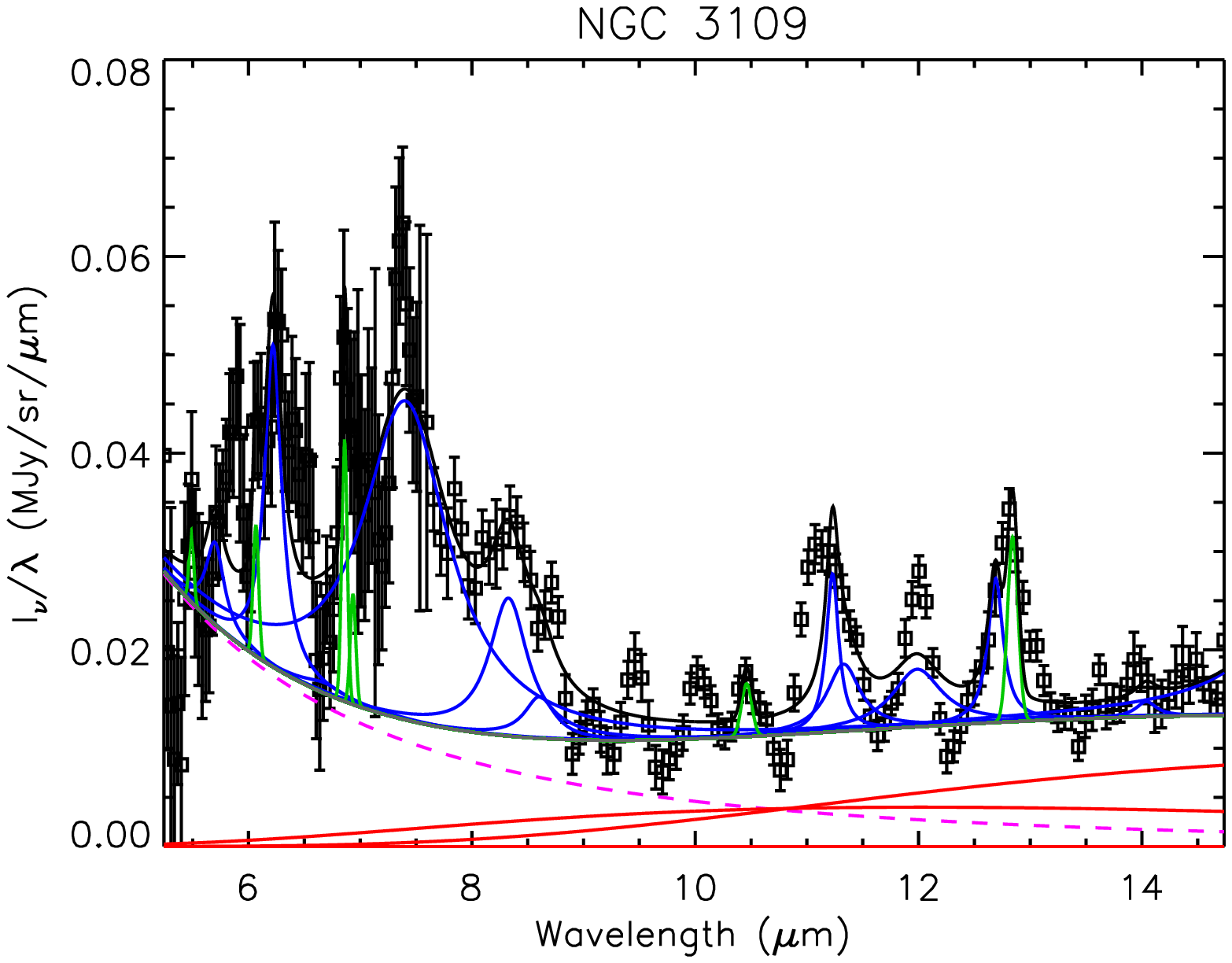}
\caption{Decomposed global IRS spectrum for NGC\,3109 produced by
  PAHFIT. The red lines represent thermal dust continuum components at
  various temperatures, the dashed magenta line the stellar continuum,
  the gray line the total (dust + stellar) continuum. The blue line
  represents the PAH features, the green line the unresolved atomic
  and molecular spectral lines, and the black line the full fitted
  model.}
\label{pahfit3109}
\end{figure}

\begin{figure}
\plotone{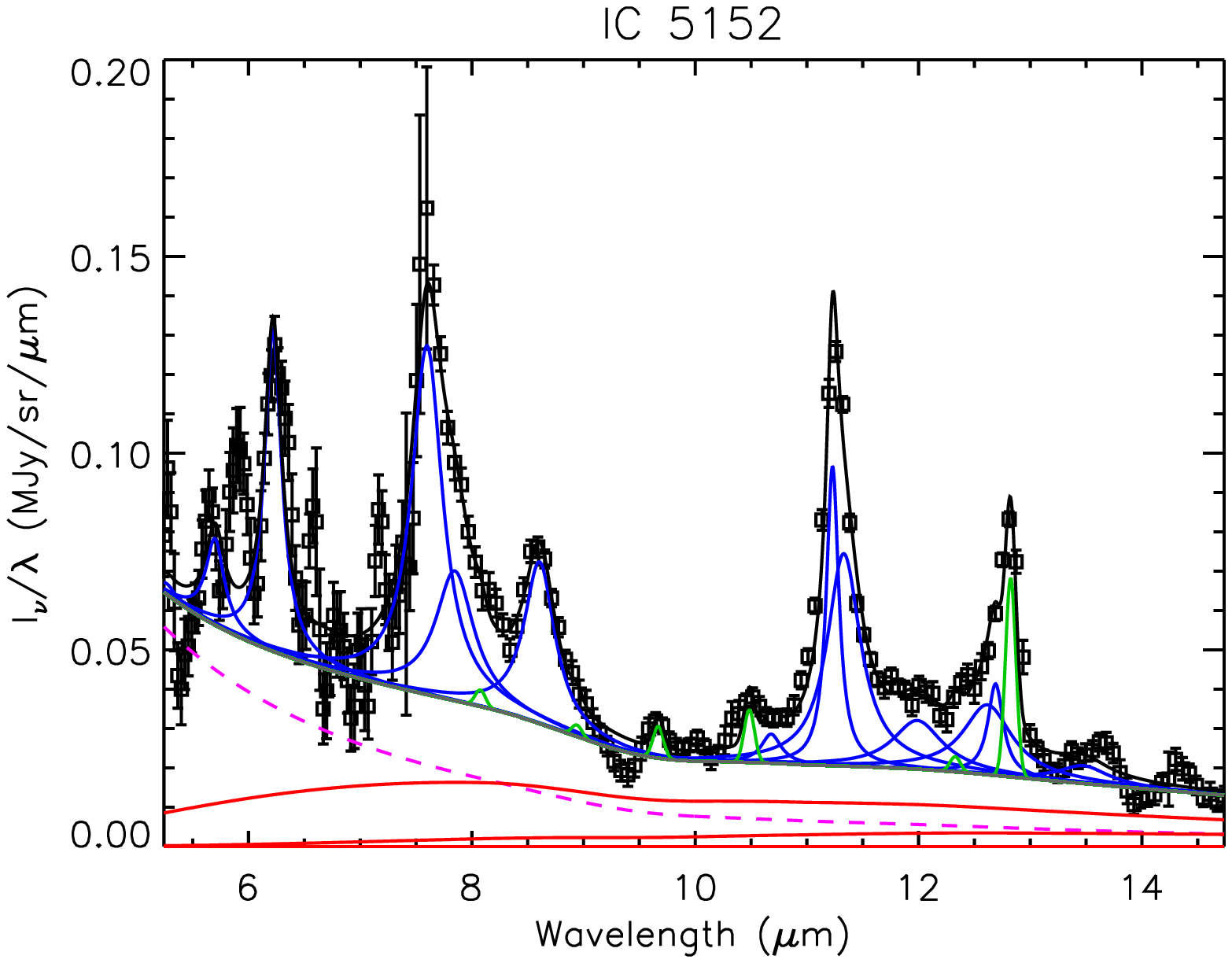}
\caption{Decomposed global IRS spectrum for IC\,5152 produced by
  PAHFIT. The red lines represent thermal dust continuum components at
  various temperatures, the dashed magenta line the stellar continuum,
  the gray line the total (dust + stellar) continuum. The blue line
  represents the PAH features, the green line the unresolved atomic
  and molecular spectral lines, and the black line the full fitted
  model.}
\label{pahfit5152}
\end{figure}

The global spectra are fitted by PAHFIT and decomposed into multiple
components, including the thermal dust continuum, stellar continuum,
PAH features, and atomic and molecular emission lines.  PAHFIT was set
such that the central wavelengths and widths of the features were not
allowed to vary between galaxies. These decomposed global spectra for
NGC\,55, NGC\,3109, and IC\,5152 are shown in Figures \ref{pahfit55},
\ref{pahfit3109}, and \ref{pahfit5152}, respectively. Multiple
components contribute to each peak visible in the full fit. For
example, using the global spectrum of NGC\,55 (Figure~\ref{pahfit55};
highest S/N) as a guide, the 7.7 $\mu$m PAH complex has contributions
from components at 7.6 $\mu$m and 7.8 $\mu$m.  Similarly, the minor
contribution of the 8.3 $\mu$m PAH feature is evident between the 8.6
$\mu$m and 7.7 $\mu$m complexes, and the 11.3 $\mu$m complex separates
into two features at 11.2 and 11.3 $\mu$m, respectively.  We remind
the reader that the PAHFIT Main Power Feature (see discussion in
\S~\ref{S2}) explicitly calculates radiances across these blended
complexes.

Figures~\ref{pahfit55}, \ref{pahfit3109}, and \ref{pahfit5152} show
that the monochromatic specific intensity at all frequencies is
highest in NGC\,55.  It is interesting to note that the stellar
continuum is weaker compared to the dust continuum in NGC\,55 than in
the other two systems; we attribute this to both a larger dust content
and a higher star formation rate in NGC\,55 than in NGC\,3109 or
IC\,5152.  We also note with interest that the fit for NGC\,3109 shows
contributions from H$_{\rm 2}$ S(5) and [Ar~II] at 6.86 $\mu$m and
6.94 $\mu$m respectively (though the errorbars are appreciable),
features that are largely absent from the other two galaxies.

The absolute values for the extracted radiances (units of
W\,m$^{-2}$\,sr$^{-1}$) of the PAH features and ionized atomic lines
are given in Table \ref{globalpahline}.  As noted above, NGC\,55 has
the largest PAH luminosity in our sample; it has $\gsim$ 5 and $\gsim$
10 times higher radiances in each PAH feature than IC\,5152 or
NGC\,3109, respectively.  NGC\,55 also has the highest ionized atomic
line radiances (compare the [Ne~II] 12.8 $\mu$m values).  Given these
properties, the appearance of the modest ionization potential [Ar~II]
line (15.76 eV; {Cox 2000}\nocite{Cox00}) in only NGC\,3109 is not
surprising.  The other two systems have stronger radiation fields: the
[S~IV] 10.5 $\mu$m feature in NGC\,55 has an ionization potential of
34.79 eV, and the [Ne~II] 12.8 $\mu$m line in both NGC\,55 and
IC\,5152 has an ionization potential of 21.56 eV.  The PAH features
overwhelm the [Ar~II] line in these systems.

We compare the ratios of the radiances of the four prominent PAH
features in a global sense for each galaxy in Table \ref{globalcomp}.
The global (8.6 $\mu$m)/(11.3 $\mu$m) ratio is the same
(0.58\,$\pm$\,0.19) for all three galaxies within the measurement
uncertainties. The (6.2 $\mu$m)/(11.3 $\mu$m), (7.7 $\mu$m)/(6.2
$\mu$m), and (7.7 $\mu$m)/(11.3 $\mu$m) ratios have more significant
scatter (factors as large as 3-4). It is not immediately clear why the
(8.6 $\mu$m)/(11.3 $\mu$m) ratio is the most stable across all three
galaxies. Considering the small sample size, it is possible that the
small dispersion is a coincidence. Since the 8.6 $\mu$m and 11.3
$\mu$m PAH features are postulated to have contributions from CH
bending modes (though the CH contribution to the 8.6 $\mu$m feature is
weak when the PAH carriers are ionized; see \S~\ref{S1.1}), the small
dispersion in those bands could be due to the ionization state of the
PAHs. However, as mentioned above, the hardness of the radiation field
in these systems varies considerably on both the global and the local
level.  Further, differences in ionization state do not explain why
statistically significant variations between other bands that arise
due to the same type of mode in similar ionization states (e.g., the
(7.7 $\mu$m)/(6.2 $\mu$m) ratio, which is dominated by the CC
stretching modes when the carriers are ionized) are observed.  The
constancy of the global (8.6 $\mu$m)/(11.3 $\mu$m) ratio is thus
likely due, at least in part, to a similar grain size distribution in
this sample.

\subsubsection{Comparing with Other Galaxies}
\label{S3.2.2}

Previous investigations have examined the variations (or lack thereof)
in PAH/PAH ratios both globally among galaxies
\citep[e.g.,][]{Smith07a,Galliano08,Hunt10} and on a spatially
resolved basis within galaxies \citep[e.g.,][]{Galliano08, Gordon08}.
From these works, evidence is mounting that the relative intensities
of the major PAH features are remarkably uniform, even across a wide
range of local and global environments. The \citet{Smith07a} and
\citet{Galliano08} samples include a wide variety of sources (spirals,
AGNs, dwarfs, H~II regions); the \citet{Hunt10} sample is composed
entirely of blue compact dwarfs (BCDs); the \citet{Gordon08} study
probes individual H~II regions in M~101.  Such a collection of
environments harbors a wide range of stellar populations, current
SFRs, metal and dust contents, and large-scale dynamical processes.
It would then be expected that the relative radiances of ionized (6.2
$\mu$m, 7.7 $\mu$m and 8.6 $\mu$m) versus neutral (11.3 $\mu$m), and
large versus small (in the same ionization state; e.g., 8.6 $\mu$m
versus 6.2 $\mu$m), PAH features would also vary considerably.

\begin{figure}
\begin{center}
\includegraphics[width=.8\textwidth,angle=90]{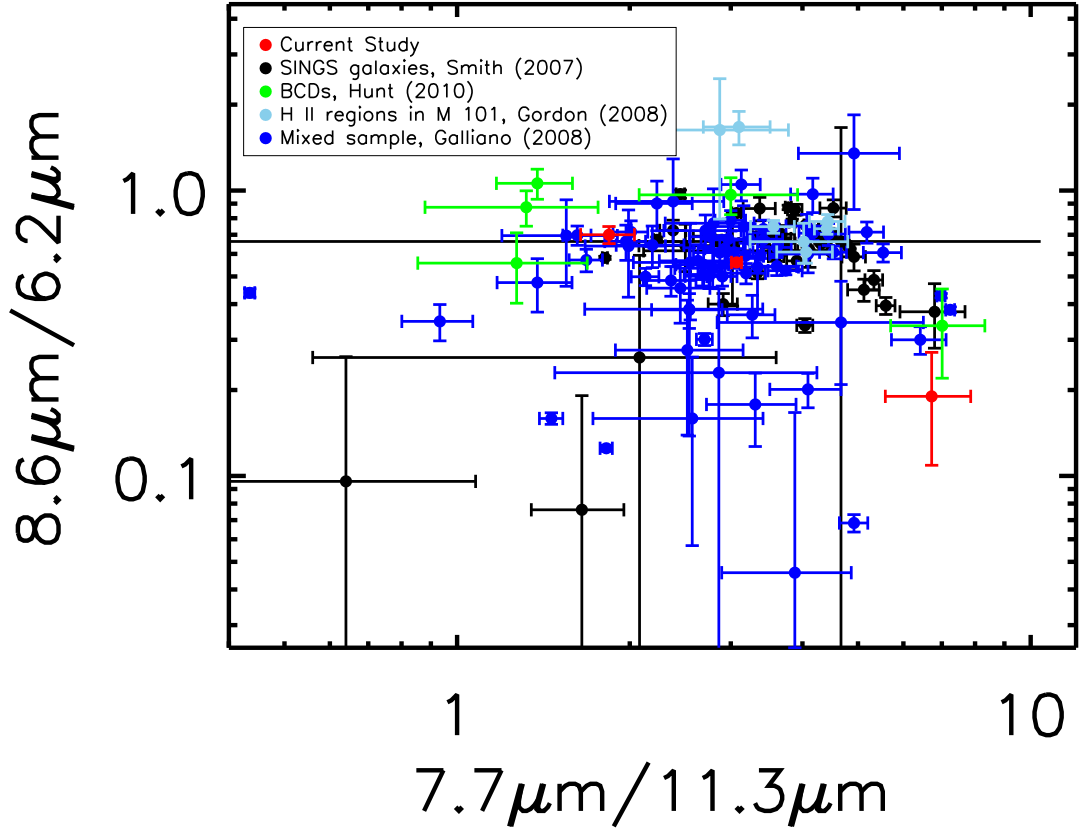}
\end{center}
\caption{Global comparison of PAH/PAH radiance ratios for individual
  galaxies and selected massive star formation regions. The four
  systems with the lowest (8.6 $\mu$m)/(6.2 $\mu$m) ratios are NGC\,4725,
  NGC\,1291, IRAS\,23128-5919, and NGC\,253, in order of increasing
  (7.7 $\mu$m)/(11.3 $\mu$m) ratio. This plot is designed to show
  contributions from larger PAH molecules increasing up the y axis,
  and contributions from ionized PAH moecules increasing along the x
  axis. The galaxies from our sample fall within the range established
  by the larger samples.}
\label{smith1}
\end{figure}

To examine both theories (grain size and ionization state) across this
broad sample of environments, we plot in Figure \ref{smith1} the three
galaxies in this sample, the inner few kiloparsecs of a subset of
SINGS galaxies from \citet{Smith07a}, a wide range of galaxy types
from \citet{Galliano08}, the sample of BCDs from \citet{Hunt10}, and
the star formation regions in M~101 studied by \citet{Gordon08}.  This
plot is designed to show contributions from larger PAH molecules
increasing up the y axis, and contributions from ionized (as compared
to neutral) PAH molecules increasing along the x axis. All data
represented were decomposed using PAHFIT, with the exception of the
data points from \citet{Galliano08}, who used their own spectral
decomposition method (in their study, {Galliano
  \etal\ 2008}\nocite{Galliano08} use two fitting methods, Spline and
Lorentzian; we present only the Lorentzian method, as it more closely
approximates the PAHFIT method).  Sources were only included on our
plot if the radiances of all four PAH features were available.  We
stress that the plot contains information about both entire galaxies
(e.g., the BCDs) and individual regions within galaxies.  Further, the
data acquisition techniques (e.g., spectral mapping versus single
pointing), analysis techniques (see above), and sensitivities may
differ from one data point to another.  While we remain mindful of
these caveats, this collection of data represents, to our knowledge,
all currently available information on the radiances of the 6.2 $\mu$m,
7.7 $\mu$m and 8.6 $\mu$m PAH features over a range of ambient
metallicities within nearby galaxies.

We find that our three galaxies nicely fit within the overall trend
established by the larger (and broader) samples, which on the whole
show a minimal range in size distribution across an order of magnitude
in ionization state (though outliers do exist; see the caption of
Figure~\ref{smith1}). This implies that a more or less constant grain
size distribution exists over the appreciable range in ionization
fraction covered by this sample.  Considering the three galaxies in
the present study (shown in red in Figure~\ref{smith1}), NGC\,55 falls
in the center of the locus of points from the larger compilation; this
is not surprising, given its comparatively large mass, metallicity,
and dust content.  IC\,5152 also falls within the larger distribution,
though at a lower ionization index than NGC\,55.  It may seem
surprising that NGC\,3109 appears at the highest ionization index
(i.e., largest global (7.7 $\mu$m)/(11.3 $\mu$m) ratio) in this plot;
however, this is easily interpreted as its comparatively low S/N
spectrum being heavily weighted toward the region of highest IR
surface brightness (and thus most intense radiation field; we return
to this point in \S~\ref{S4}).

Taken as a whole, Figure~\ref{smith1} is evidence for a similar PAH
size distribution in spirals, AGN, dwarfs, and H~II regions.  This
suggests that the emission properties of the PAH carriers are more
significantly affected by their local environment than by the past
growth/destruction history (in agreement with previous works, e.g.,
{Jackson \etal\ 2006}\nocite{Jackson06}).  It also indicates that the
growth/destruction history is not measurably affected by local
effects, and that instead the grain size distribution, if not varying
or affected on the local level, is instead affected by larger scale,
more long-term processes. It is also apparent that dwarf galaxies
constitute some of the most extreme outliers from the general trend
seen in Figure~\ref{smith1}: the BCDs from the \citet{Hunt10} sample
(shown in green) appear at the extrema of the locus of points from the
\citet{Smith07a} and \citet{Galliano08} studies, suggesting a wide
range in global ionization indices in these dwarfs.

\begin{figure}
\begin{center}
\includegraphics[width=.8\textwidth]{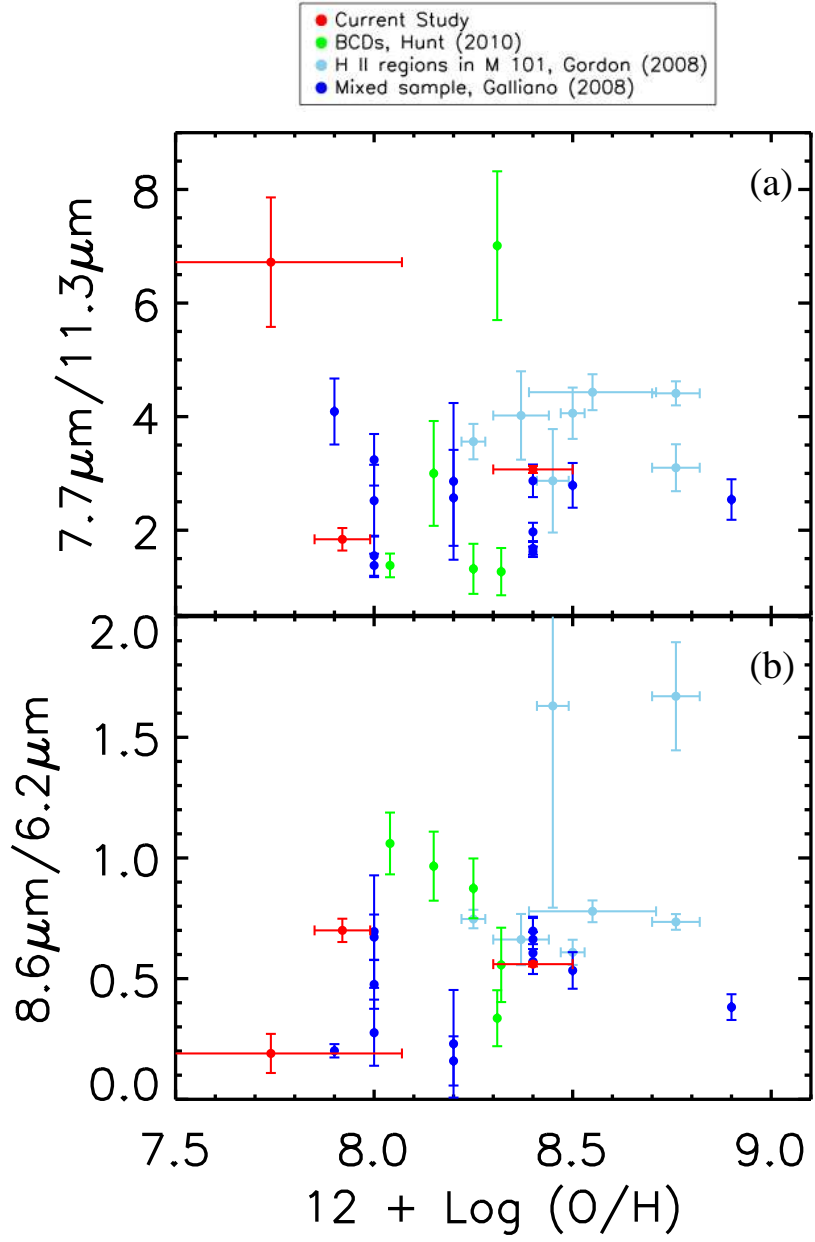}
\end{center}
\caption{Comparison of PAH/PAH radiance ratios with gas-phase oxygen
  abundances in dwarf galaxies and in individual H~II regions in
  M~101. No uncertainties in metallicity were available for the
  \citet{Hunt10} or \citet{Galliano08} data. Only very weak trends
  with metallicity are apparent from this plot.}
\label{metals}
\end{figure}

One may be tempted to ascribe the extremal ratios found in some of
(but not all) the dwarf galaxies to metallicity effects.  To examine
this possibility, we plot in Figure~\ref{metals} the global
(8.6 $\mu$m)/(6.2 $\mu$m) and (7.7 $\mu$m)/(11.3 $\mu$m) ratios versus gas-phase
oxygen abundances for a subset of the systems in Figure~\ref{smith1}.
In creating this plot, we implicitly assume no abundance gradients in
the dwarfs, and we explicitly exclude global values for spirals, as
they have well-documented abundance gradients.  We do, however,
include the measurements of individual H~II regions in M~101 (each
with abundance measurements) presented by \citet{Gordon08}. In
interpreting this plot, we remain mindful that the dataset is
inhomogeneous in the sense that the local conditions within individual
star formation regions of a spiral disk will be quite distinct from
those within a dwarf galaxy.  We also stress again that the high
global (7.7 $\mu$m)/(11.3 $\mu$m) ratio in NGC\,3109 is heavily weighted
toward the region of highest IR surface brightness.  With these
caveats in mind, we find only very weak evidence for a trend of
increasing grain size or ionization index with increasing metal
abundance.  This is in agreement with previous studies
\citep[e.g.,][]{Jackson06,Rosenberg06,Gordon08}, where the effect of
metallicity has been found to be secondary to other factors in
controlling the intensities of PAH emission features.  While the total
amount of metals in a galaxy is clearly an important property in
governing the intensity of PAH emission
\citep{Engelbracht05,Engelbracht08}, evidence shows that the relative
intensities of the PAH features show a remarkable degree of uniformity
in the systems considered here.

\section{Spatially Resolved Emission}
\label{S4}

\begin{figure}
\begin{center}
\includegraphics[width=.9\textwidth]{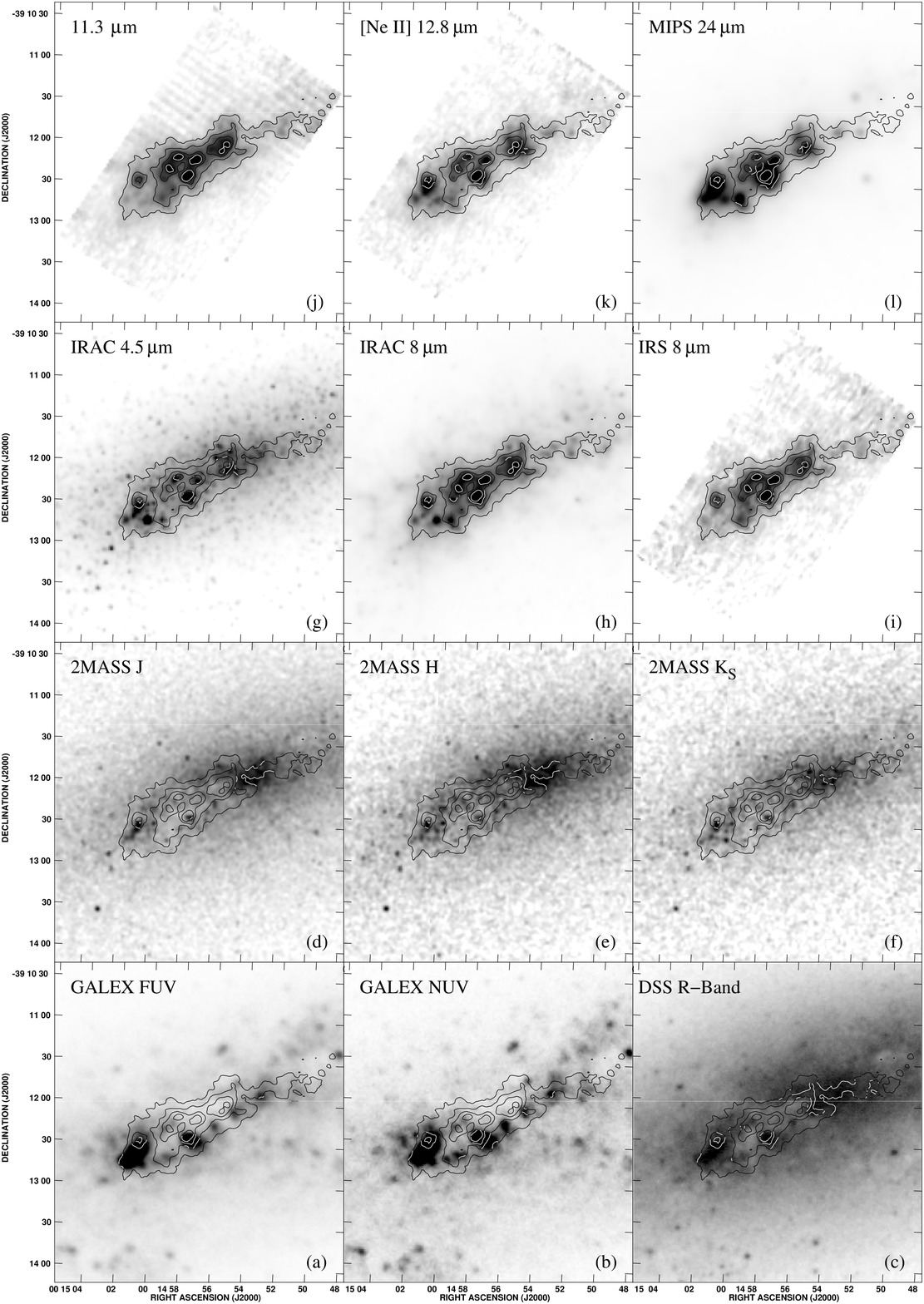}
\end{center}
\caption{Multiwavelength images of the region of NGC\,55 covered by
  the IRS spectral map, ordered by increasing wavelength from (a) to
  (l) as labeled.  Contours of 11.3 $\mu$m surface brightness are
  overlaid in each panel at the 1.5, 3, 4.5 and 6 MJy\,sr$^{-1}$
  levels.}
\label{contour55}
\end{figure}

\begin{figure}
\begin{center}
\includegraphics[width=.99\textwidth]{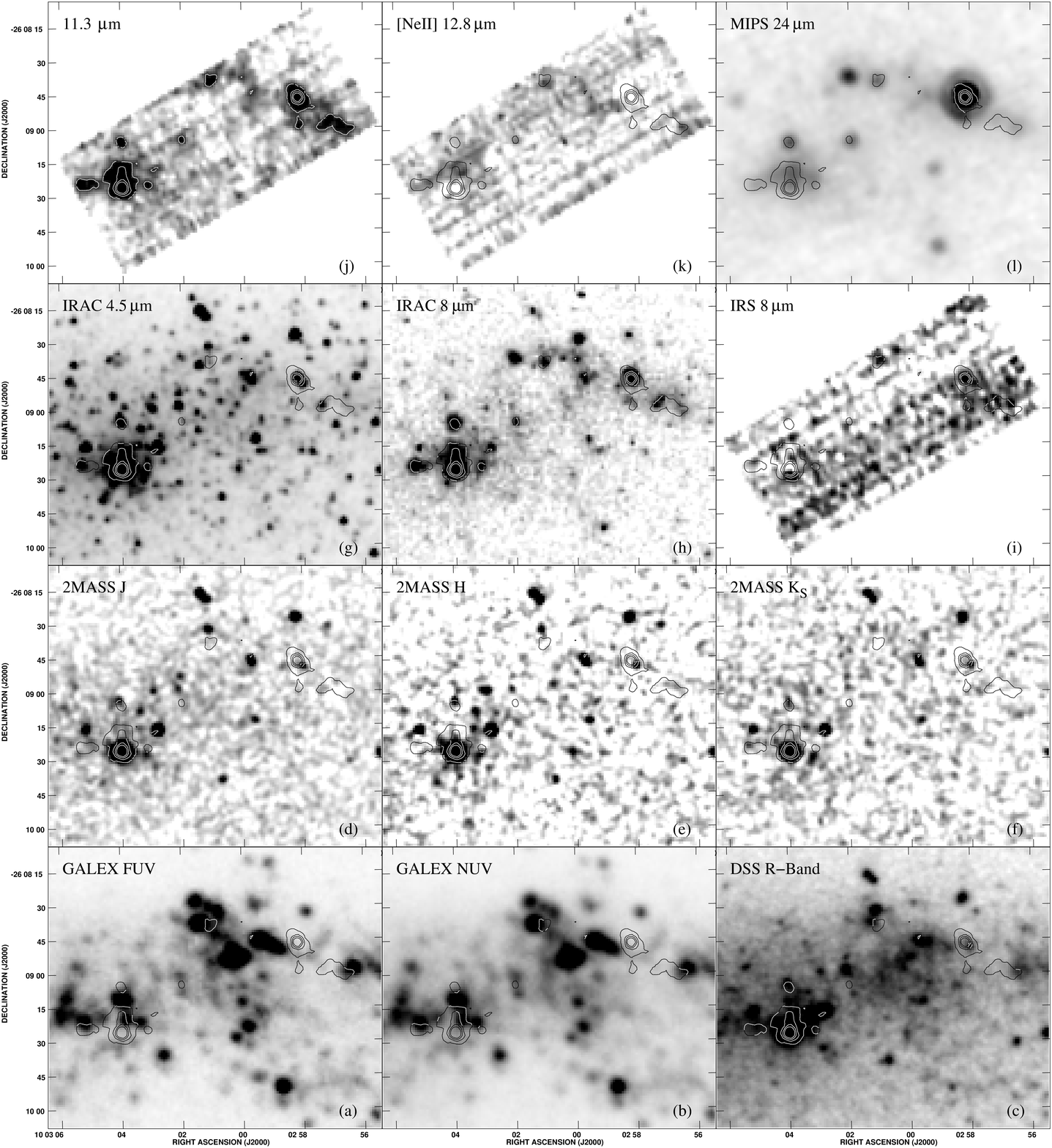}
\end{center}
\caption{Multiwavelength images of the region of NGC\,3109 covered by
  the IRS spectral map, ordered by increasing wavelength from (a) to
  (l) as labeled.  Contours of 11.3 $\mu$m surface brightness are
  overlaid in each panel at the 0.5, 1.0, 1.5 and 2.0 MJy\,sr$^{-1}$
  levels.}
\label{contour3109}
\end{figure}

\begin{figure}
\begin{center}
\includegraphics[width=.95\textwidth]{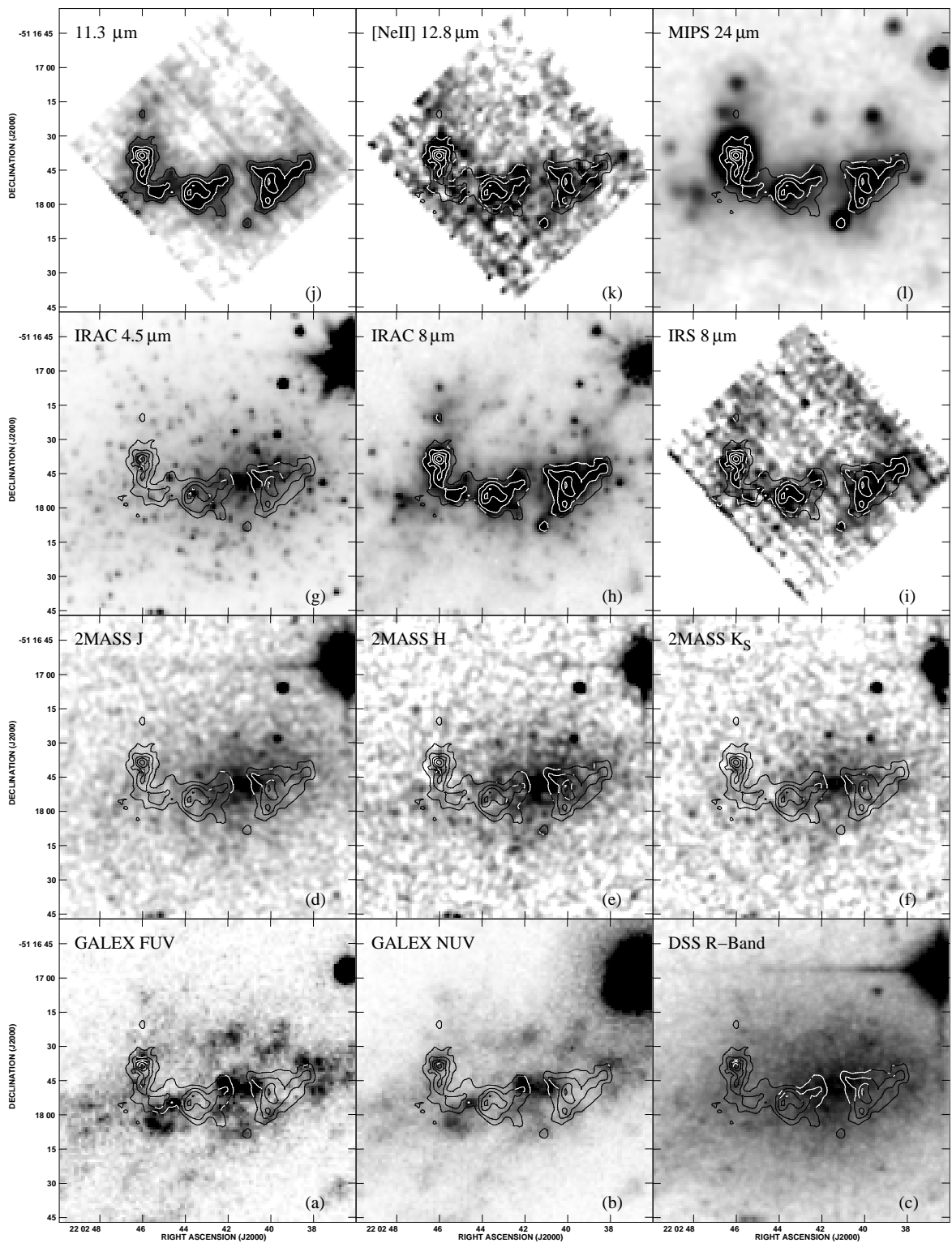}
\end{center}
\caption{Multiwavelength images of the region of IC\,5152 covered by
  the IRS spectral map, ordered by increasing wavelength from (a) to
  (l) as labeled.  Contours of 11.3 $\mu$m surface brightness are
  overlaid in each panel at the 1.0, 1.5, 2.0, 2.5 and 3.0
  MJy\,sr$^{-1}$ levels.  The extremely luminous source in the upper
  right is a Galactic foreground star.}
\label{contour5152}
\end{figure}

In order to examine the comparative spatial distributions of dust,
gas, stellar populations, and radiation field strengths in our sample
galaxies, images are shown at various wavelengths in Figures
\ref{contour55}, \ref{contour3109}, and \ref{contour5152} for NGC 55,
NGC\,3109, and IC\,5152, respectively.  Here we exploit our mapping
observational strategy (see \S~\ref{S2}).  Using the global spectrum
of each galaxy, individual spectral regions are selected that bracket
PAH or emission lines of interest.  The underlying continuum is
removed using areas of the spectrum outside the feature (we note the
difficulty of accurate continuum subtraction for some of the wide PAH
features, especially the blended features in the 7 $\mu$m -- 9 $\mu$m
range).  The spatial location of the regions that contribute are
known; a map is then constructed of the emission in a given wavelength
interval (see {Smith \etal\ 2007b}\nocite{Smith07b} for details).  The
complexity of the spectrum of each system (see
Figures~\ref{idlglobalspec55}, \ref{idlglobalspec3109}, and
\ref{idlglobalspec5152}) requires fine-tuning of the spectral
extraction regions for each feature individually.  We explicitly list
the wavelength intervals over which the emission maps are created for
each system in Table~\ref{extract}.

Panels (i), (j), and (k) of Figures~\ref{contour55},
\ref{contour3109}, and \ref{contour5152} show images created from the
IRS data.  The spatial distributions of IRS 11.3 $\mu$m (see panel j
of each figure) and broad 8 $\mu$m (integrated from 7.288 $\mu$m --
8.900 $\mu$m in all systems; note that this includes contributions
from both the 7.7 $\mu$m and the 8.6 $\mu$m PAH complexes; see panel i
of each figure) emission correlate well in all three galaxies. The
IRAC and IRS 8 $\mu$m maps also agree in most areas; the small
observed differences are most likely due to the larger bandwidth of
the IRAC 8 $\mu$m filter (extending from $\sim$6.4 $\mu$m -- $\sim$9.4
$\mu$m).  As a check on the quality of the absolute flux calibration
of the IRS spectra, we explicitly compared the total flux density in
the IRAC 8 $\mu$m band with the total flux density integrated over the
matching IRAC bandpass in the IRS spectrum of each galaxy; while these
images are not shown in Figures~\ref{contour55}--\ref{contour5152},
the resulting flux densities agree at the 8\%, 13\%, and 17\% levels
for NGC\,55, IC\,5152, and NGC\,3109, respectively.

There is very good spatial correlation between emission from PAH
features and ionized gas as indicated by [Ne~II] emission (see panel k
of each figure) in NGC\,55 and IC\,5152.  This immediately suggests
that the PAH emission is arising from regions with comparatively hard
radiation fields (recall that the ionization potential of [Ne~II] is
$\sim$21.6 eV) in these systems.  There is a much poorer spatial
agreement between [Ne~II] and PAH emission in NGC\,3109.

It is interesting to note that in only a few regions in these galaxies
do areas of high surface brightness UV emission coincide with areas of
high surface brightness PAH emission. There are locations in each
system that are UV-bright and PAH-dim (arguing for minimal extinction)
as well as regions that have high PAH surface brightnesses but little
observed UV or optical emission (arguing for pronounced local
extinction).  Examining each system from the UV to 4.5 $\mu$m in
Figures~\ref{contour55}, \ref{contour3109} and \ref{contour5152}, we
conclude that the multiwavelength characteristics of NGC\,55 and
NGC\,3109 are consistent with variable extinction of up to a few
magnitudes in the UV; both systems show good agreement between 11.3
$\mu$m PAH surface brightness and 4.5 $\mu$m morphology.  IC\,5152 is
more challenging to interpret: its near-infrared stellar population
bears a strikingly different morphology than the PAH or [Ne~II]
emission.  This can be interpreted as more severe localized extinction
than in NGC\,55 or NGC\,3109; such variations have been observed in
dwarf galaxies in previous works \citep[e.g.,][]{Cannon06a}.  These
implied optical and UV extinctions are in agreement with the global
values derived by \citet{Lee09}, albeit with significant internal
variations.

\begin{figure}
\plotone{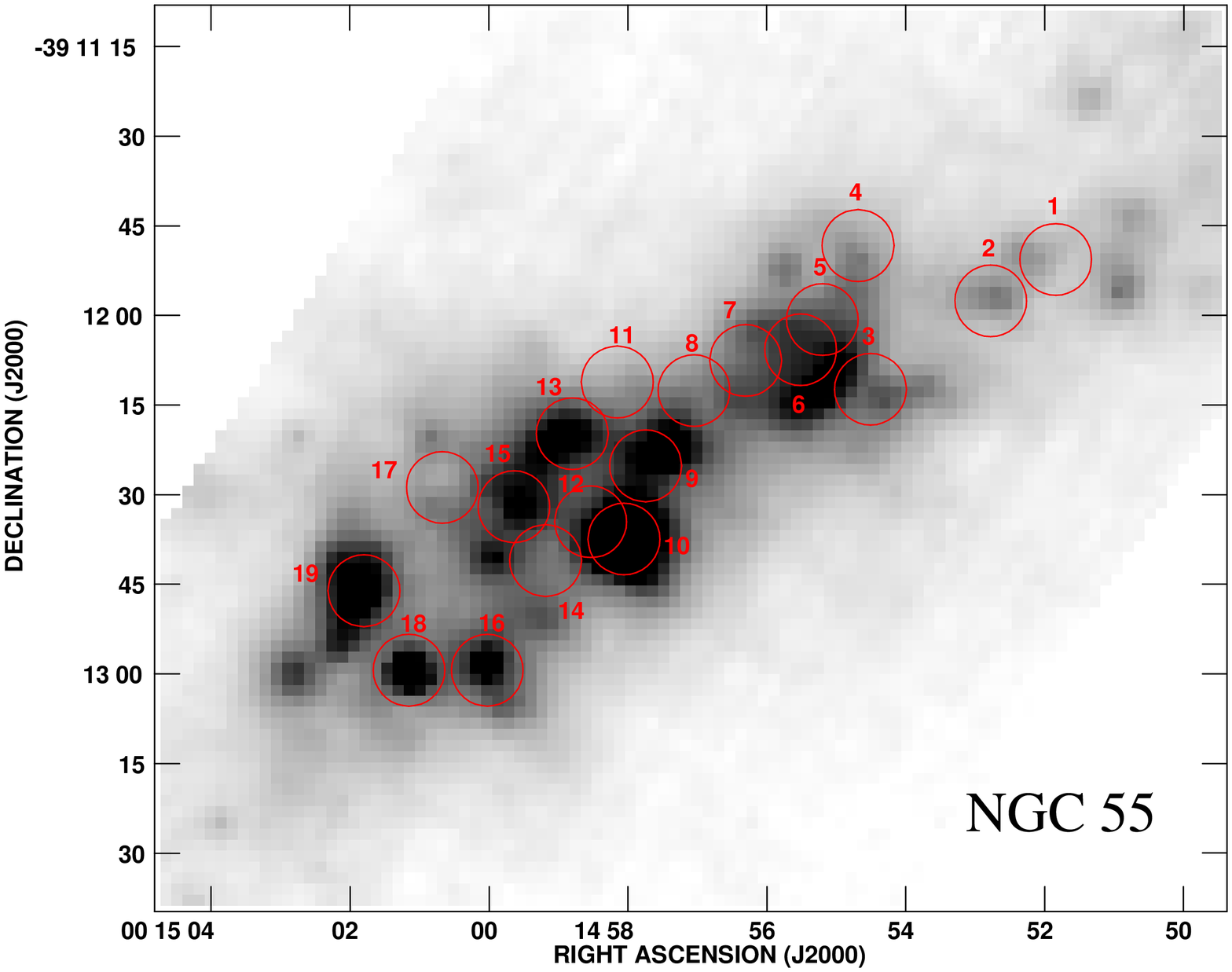}
\caption{Full, integrated IRS map of NGC\,55 with extraction regions
  labeled. Regions have a physical radius of 52.5 pc.}
\label{region55}
\end{figure}

\begin{figure}
\plotone{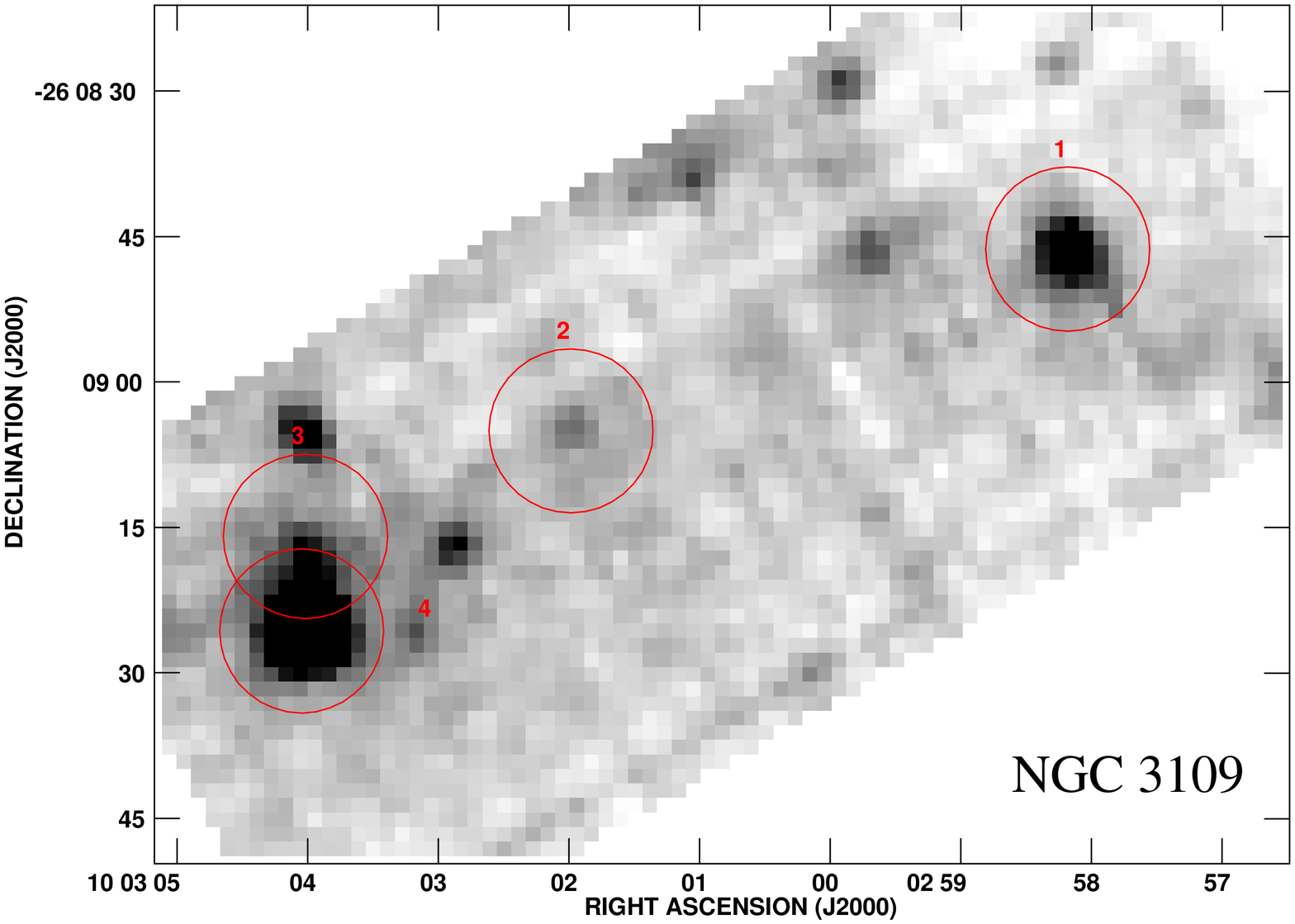}
\caption{Full, integrated IRS maps of NGC\,3109 with extraction regions
  labeled. Regions have a physical radius of 52.5 pc.}
\label{region3109}
\end{figure}

\begin{figure}
\plotone{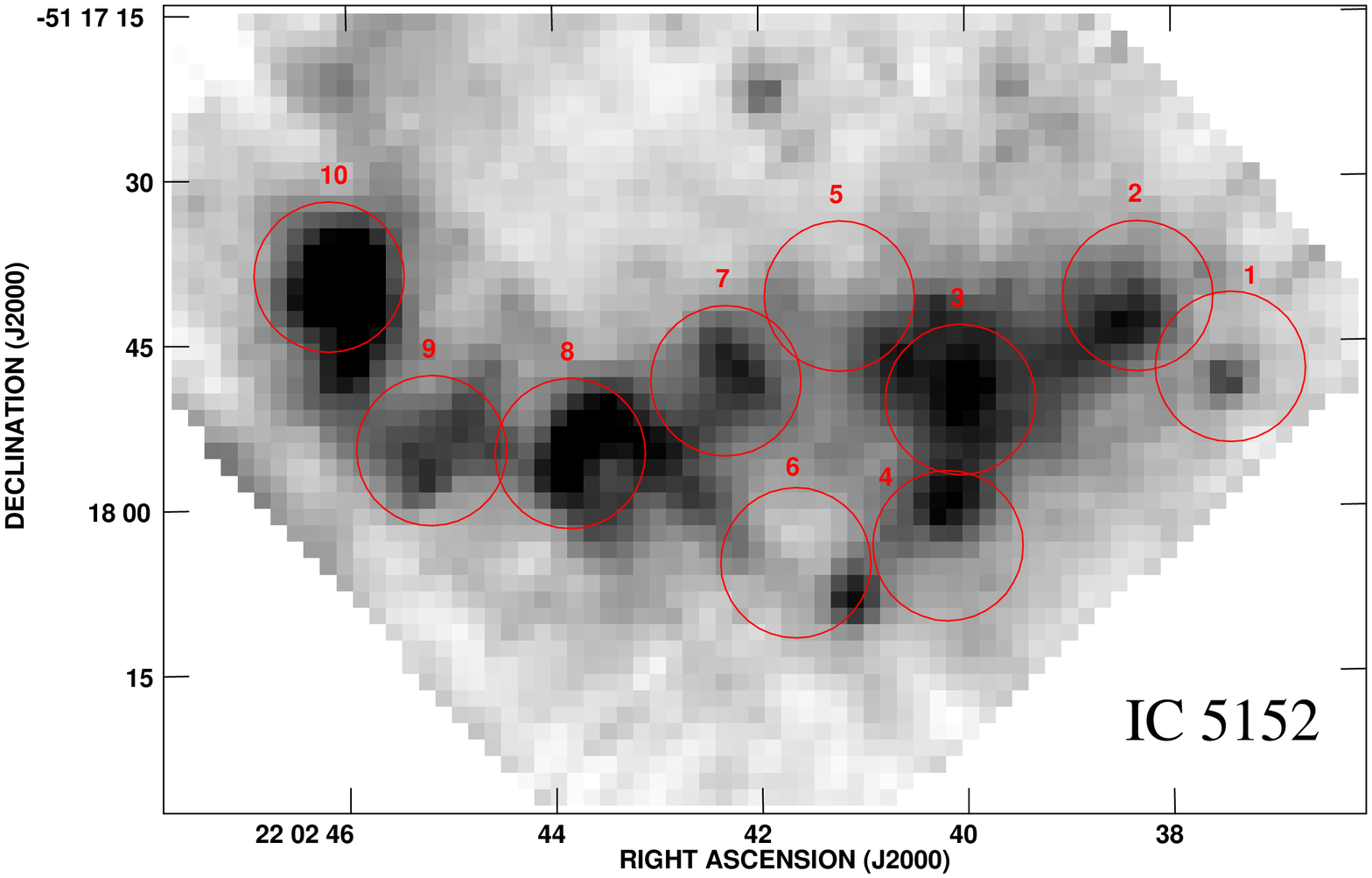}
\caption{Full, integrated IRS map of IC\,5152 with extraction regions
  labeled. Regions have a physical radius of 52.5 pc.}
\label{region5152}
\end{figure}

In all three galaxies, we examine PAH ratios for individual sources
that are spatially resolved in our spectral image data cubes. Spectra
were extracted from CUBISM maps using circular regions of 52.5 pc
radius. These extraction regions can be seen in Figures \ref{region55}
(NGC 55), \ref{region3109} (NGC\,3109), and \ref{region5152} (IC
5152). 52.5 pc is four times the smallest spatial resolution possible
with the data, or four pixels wide on the IRS map of NGC\,55 (D = 2.17
Mpc, the most distant system). At 8.6 $\mu$m, {\it Spitzer}'s
diffraction limit is 2.55\arcsec; 52.5 pc corresponds to
5.0\arcsec\ at the distance of the NGC\,55. For NGC\,3109 and
IC\,5152, the extraction regions are 8.1\arcsec\ and 5.5\arcsec,
respectively. Extraction regions were chosen to be the same physical
size in each galaxy for optimal comparison of features. 52.5 pc radius
regions were judged to best fit physical variations in structure
across the varying angular size of all three galaxies; further, 52.5
pc is large enough to analyze on NGC 55, but small enough to allow
multiple regions to be extracted from NGC\,3109 and IC\,5152. The
extracted spectrum of each region was then fitted with PAHFIT to
derive PAH feature and emission line radiances. The results of these
fits are shown in Tables~\ref{ratios55_1}, \ref{ratios55_2},
\ref{ratios3109_1}, \ref{ratios3109_2}, \ref{ratios5152_1}, and
\ref{ratios5152_2}, and discussed in detail in \S~\ref{S4.1},
\S~\ref{S4.2}, and \S~\ref{S4.3} below.

\begin{figure}
\begin{center}
\includegraphics[width=.95\textwidth,angle=90]{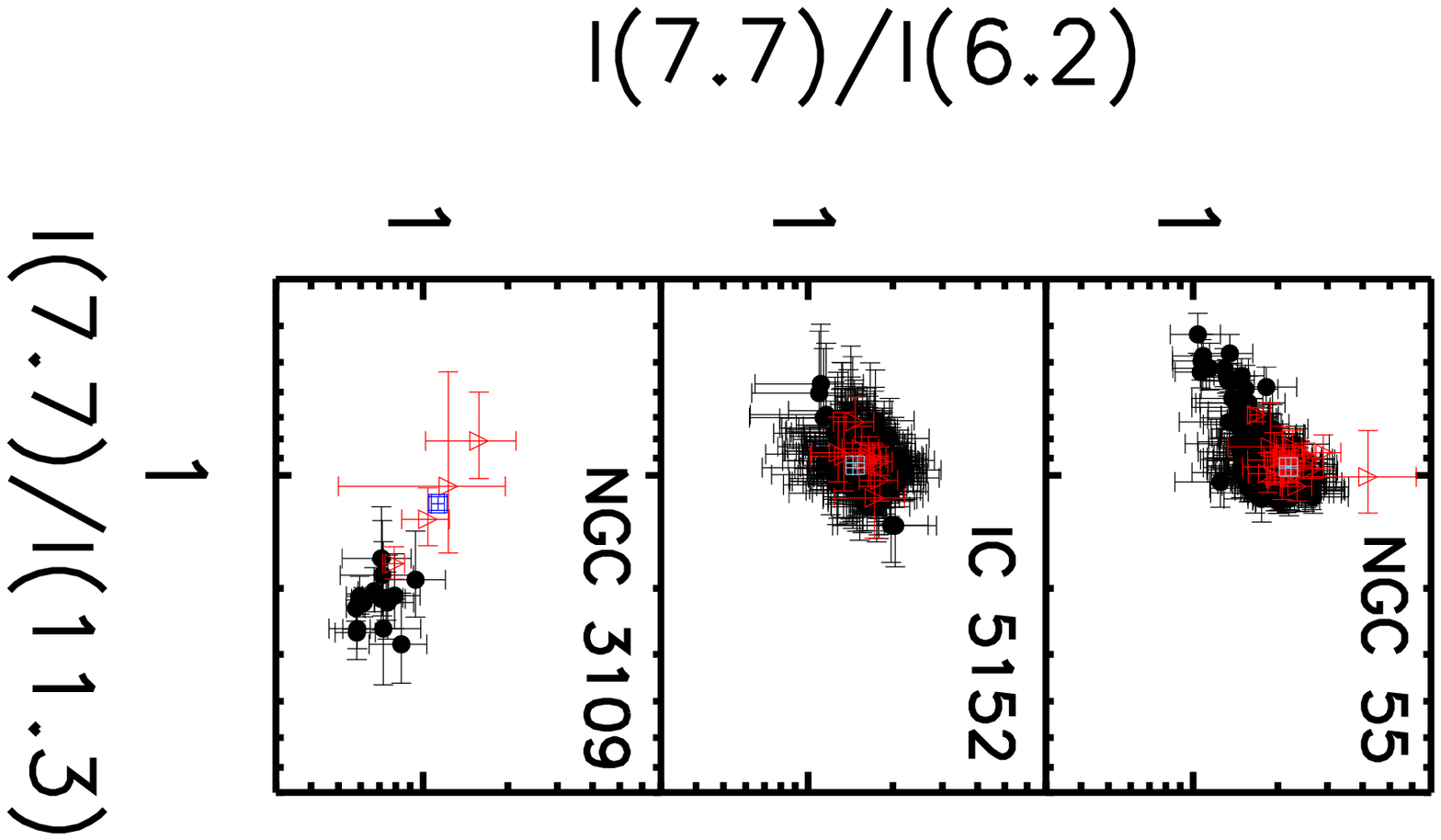}
\end{center}
\caption{PAH radiance ratios within each galaxy. Pixel by pixel
  comparisons are shown in black, trimmed to show data above the
  3$\sigma$ level (note that in NGC\,3109, which has generally low S/N,
  this causes an offset between the pixel cloud and the integrated
  regions, which include all pixels regardless of S/N level; see
  discussion in \S~\ref{S4}). Points in red represent the integrated
  values from the 52.5 pc extracted regions. The blue point represents
  the global ratio.}
\label{gall1}
\end{figure}

\begin{figure}
\begin{center}
     \includegraphics[width=.95\textwidth,angle=90]{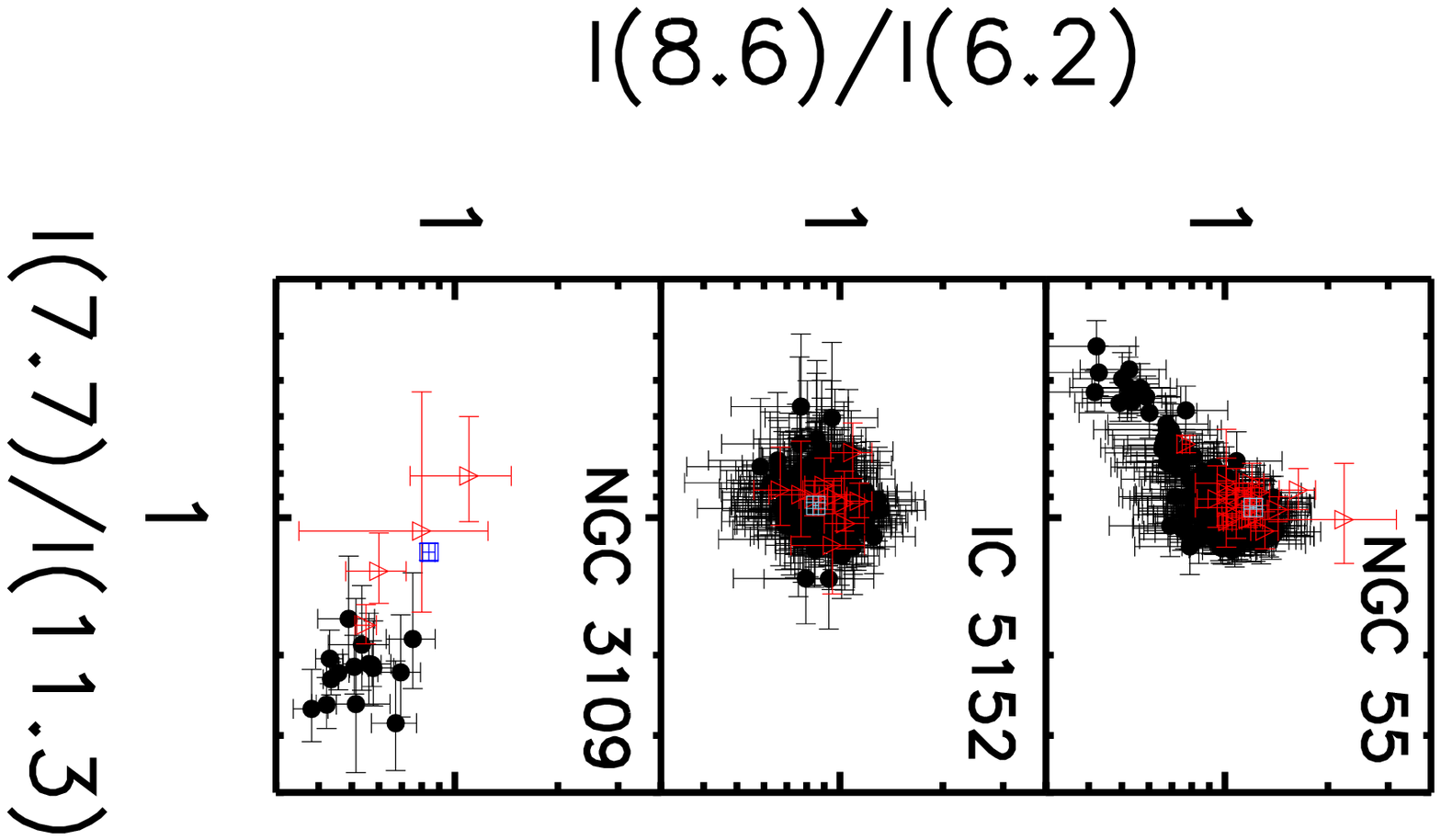}
\end{center}
\caption{PAH radiance ratios within each galaxy. Pixel by pixel
  comparisons are shown in black, trimmed to show data above the
  3$\sigma$ level (note that in NGC\,3109, which has generally low S/N,
  this causes an offset between the pixel cloud and the integrated
  regions, which include all pixels regardless of S/N level; see
  discussion in \S~\ref{S4}). Points in red represent the integrated
  values from the 52.5 pc extracted regions. The blue point represents
  the global ratio.}
\label{gall2}
\end{figure}

PAH band variations are also examined on a pixel by pixel basis; we
show these results in Figures~\ref{gall1} and \ref{gall2}. The data
included in these plots were not calculated by PAHFIT. Instead,
spectral maps were extracted from CUBISM; individual pixel specific
intensities, integrated specific intensities within the apertures
described above (see Figures~\ref{region55}, \ref{region3109}, and
\ref{region5152}), and integrated global specific intensities were
then derived from these images.  The reported errors are taken
directly from the error maps as produced by CUBISM.  Only pixels with
a S/N ratio of 3 or higher are included in the pixel-by-pixel
analysis; all pixels in a given region are included for the region and
global values.

Figures~ \ref{gall1} and \ref{gall2} show that the
(7.7 $\mu$m)/(11.3 $\mu$m) PAH radiance ratios span only a factor of $\sim$5
within each of these galaxies.  Interestingly, for NGC\,55 and
IC\,5152, the regions are in good agreement with the pixel
data. However, for NGC\,3109, a marked shift to lower
(7.7 $\mu$m)/(11.3 $\mu$m) ratios is seen for the integrated regions
compared to the individual pixel values.  This is due to the
uncertainty cut on the pixels, which removes low S/N pixels from the
data. This means that the extraction regions, which had no uncertainty
cut applied, have the potential to contain pixels of a much lower
specific intensity than is possible in the individual pixel
data. Conversely, the individual pixels are therefore weighted towards
higher surface brightness regions of the galaxy in a way the
extraction regions are not. This difference is most notable in low S/N
regions like Region 2. Region 4, the highest surface brightness region
in NGC\,3109, is well within the average range of the rest of the
pixels. Examination of Figure \ref{contour3109} shows more diffuse
11.3 $\mu$m emission than other bands in NGC\,3109, including the IRS
8 $\mu$m band, which encompasses the 7.7 $\mu$m PAH feature. From this
we conclude that integrating over lower surface brightness regions
would indeed produce a larger contribution from 11.3 $\mu$m compared
to 7.7 $\mu$m emission, leading to a smaller ratio of these two
values.

The small changes (a factor of $\sim$5, or just larger than the
associated uncertainties for most regions) in the (7.7 $\mu$m)/(11.3 $\mu$m)
PAH radiance ratios can be interpreted as a relatively minor variation
of the ionization index (i.e., radiation field strength) within a
given galaxy.  For NGC\,55 and IC\,5152, this is to be expected from
examination of Figures~\ref{contour55} and \ref{contour5152}.  The
agreement in morphology between the [Ne~II] emission line and the IRS
8 $\mu$m and 11.3 $\mu$m PAH bands shows that the latter two are
primarily arising in regions rich in energetic photons.  The
interpretation of the NGC\,3109 plot is more ambiguous, primarily
because there are so many fewer pixels that contribute above the
3$\sigma$ level compared to the two more massive systems.  However, at
face value, the variation of the (7.7 $\mu$m)/(11.3 $\mu$m) PAH radiance
ratio across NGC\,3109 implies that the radiation field strength is
high in one specific region (where the individual pixels achieve a S/N
ratio above 3) and lower throughout most of the system.  This is also
apparent by examining Figure~\ref{contour3109}.

Two diagnostics of the PAH size distribution are plotted against the
(7.7 $\mu$m)/(11.3 $\mu$m) PAH radiance ratio in Figures~\ref{gall1} and
\ref{gall2}, the former via the (7.7 $\mu$m)/(6.2 $\mu$m) PAH radiance ratio
and the latter via the (8.6 $\mu$m)/(6.2 $\mu$m) PAH radiance ratio.  Note
that both of these ratios are dominated by the CC stretching modes in
regions where the PAH carriers are ionized (see the distributions of
[Ne~II] emission throughout each system in Figures~\ref{contour55},
\ref{contour3109} and \ref{contour5152}). Recall that the 7.7 $\mu$m
PAH feature arises from a mixed population of large and small
carriers, so the interpretation of the (7.7 $\mu$m)/(6.2 $\mu$m) ratio as
sensitive only to the PAH size distribution should be treated with
more caution than the (8.6 $\mu$m)/(6.2 $\mu$m) ratio.  From these plots we
see that NGC\,55 shows a trend of larger grain sizes (i.e., larger
ordinate values) in regions of harder radiation field (larger
abscissa values), with a more transparent correlation using the larger
difference in grain size (i.e., in the (8.6 $\mu$m)/(6.2 $\mu$m) plot).
IC\,5152 shows a slight trend in the same sense but the signature is
weaker than in NGC\,55.  NGC\,3109 shows a weak trend in the opposite
sense.  We again stress that the region and global radiances in these
figures have no minimum specific intensity threshold; hence the
reality of the trend in NGC\,3109 should be interpreted with caution.

In summary, Figures~\ref{gall1} and \ref{gall2} show variations of
radiation field strength and PAH carrier size distributions on a
spatially resolved basis.  While very weak trends can be inferred in
these data, they are nonetheless consistent with an interpretation of
very little variation in grain size distribution as a function of
radiation field strength.  This would seem to indicate that the
distribution of PAH grain sizes are long lived, and not significantly
affected by local phenomena. Stated differently, the constancy of the
PAH/PAH ratios seen amongst galaxies (see detailed discussion in
\S~\ref{S3.2} above) appears to extend down to the spatially resolved
level within individual galaxies as well.  While our study has
pre-selected regions that are IR-bright within these dwarfs (and thus
potentially having homogeneous characteristics), the variations of PAH
feature radiances on these spatial scales are very similar to the
variations seen on global scales.

\subsection{NGC\,55}
\label{S4.1}

Data for nineteen circular regions with a radius of 52.5 pc (some of
them unavoidably containing areas of overlap) were extracted across
the observed region of the galaxy (see Figure~\ref{region55}), and
their extracted radiances are shown in Table \ref{ratios55_1}. As
expected based on the results discussed above and shown in
Figures~\ref{gall1} and \ref{gall2}, we find very few statistically
significant deviations of PAH/PAH radiance ratios derived in these
apertures throughout NGC\,55.  Table~\ref{ratios55_1} shows the
weighted mean values of the (6.2 $\mu$m)/(11.3 $\mu$m),
(7.7 $\mu$m)/(11.3 $\mu$m), (8.6 $\mu$m)/(11.3 $\mu$m), (7.7 $\mu$m)/(6.2 $\mu$m) and
(8.6 $\mu$m)/(6.2 $\mu$m) radiance ratios and the standard deviation across
this sample.  Of the 95 ratios given in the table, we identify 11
ratios that deviate at more than the 2$\sigma$ level; no variations at
the 3$\sigma$ level or larger are identified.  Only two regions are
identified as having variations in more than one band: Region\,2 has
$\sim$2$\sigma$ variations in all ratios except (6.2 $\mu$m)/(11.3 $\mu$m),
while Region\,16 has $\sim$2$\sigma$ variations in the
(7.7 $\mu$m)/(11.3 $\mu$m)m, (8.6 $\mu$m)/(11.3 $\mu$m), and (7.7 $\mu$m)/(6.2 $\mu$m)
radiance ratios.

We compare the radiances of PAH features with emission lines and
broad-band MIPS\,24 $\mu$m dust continuum in Table~\ref{ratios55_2}.
The radiance of the [Ne~II] emission line compared to that of either a
neutral (11.3 $\mu$m) or an ionized (8.6 $\mu$m) PAH emission feature
show very few statistically significant variations within NGC\,55.
Similarly, the ratio of 24 $\mu$m to PAH radiances is fairly uniform
across this galaxy, though the standard deviation is much larger.

We draw attention to some interesting complexities in the data in
Table~\ref{ratios55_2}.  First, Region\,10 is extremely bright
throughout the MIR, and especially so at 24 $\mu$m (compare
Figures~\ref{contour55} and \ref{region55}); it has the largest
(24 $\mu$m)/(11.3 $\mu$m) ratio (a $>$3$\sigma$ deviation from the average)
and one of the highest ratios of 24 $\mu$m radiance to the 8.6 $\mu$m
PAH band.  It is luminous in the UV and coincident with a high-surface
brightness optical cluster. However, its ratio of [Ne~II] to PAH
radiance is average.  Region\,16 is also luminous at 24 $\mu$m, but
contains no UV or optical counterpart and again has average [Ne~II]
emission line ratios compared to the PAH features.  Finally,
Region\,19 is luminous at all wavelengths (see Figure~\ref{contour55})
and has the largest [Ne~II] to PAH ratio in NGC\,55.  Thus, although
the conditions within individual regions in NGC\,55 are diverse, we
find no statistically significant variations of PAH/PAH radiance
ratios in the observed regions of NGC\,55.

\subsection{NGC\,3109}
\label{S4.2}

We examine four 52.5 pc radius apertures in NGC\,3109.  The low S/N
ratio throughout much of the observed region limits our exploration to
only those areas that are comparatively IR-bright (see
Figure~\ref{contour3109}).  Negligible scatter is seen in the PAH/PAH
radiance ratios for these regions, as seen in Table
\ref{ratios3109_1}; only one aperture deviates from the weighted mean
at the 2$\sigma$ significance level (the (6.2 $\mu$m)/(11.3 $\mu$m) ratio
for Region\,4).  This is in marked contrast to the very significant
variations seen in the ratio of 24 $\mu$m to PAH radiance in these
four regions of the galaxy.  While the weighted mean in
Table~\ref{ratios3109_2} is affected by the difference in the
errorbars over the small number of apertures, it is clear that
Regions\,1 and 4 have very different properties.

An examination of Figure~\ref{contour3109} highlights curious
differences between these two regions.  Region\,1 contains very weak
[Ne~II] 12.8 $\mu$m emission; it clearly harbors an embedded IR
source, since it is has no associated UV emission, is fairly weak in
the IRAC 8 $\mu$m band, but very luminous at 24 $\mu$m.  This suggests
a cool thermal dust component at this location.  Taken together, these
properties suggest that the radiation field strength is low at the
location of Region\,1 and that the PAHs present have a significant
neutral component.

Regions\,3 and 4 are in close physical proximity and present an
interesting local environment with significant variations over small
physical scales.  Region\,4 is very bright in the IRAC 8 $\mu$m band
and is coincident with a high surface brightness optical cluster.
Region\,3 is much fainter at 8 $\mu$m, yet it has the largest
[Ne~II]/(PAH) ratios of any of the regions in NGC\,3109.

Despite these dramatic changes in environmental properties, the ratios
of PAH features presented in Table~\ref{ratios3109_1} are
statistically indistinguishable.  Even a cursory examination of the
panels in Figure~\ref{contour3109} highlights the complexity of the
relationship between PAH emission and local properties in galaxies
(e.g., slope of the local infrared spectral energy distribution (SED),
UV intensity, infrared emission line radiance, etc.).  The data for
NGC\,3109, while only available for selected regions, provides support
for the hypothesis of constant PAH ratios within dwarf galaxies.

\subsection{IC\,5152}
\label{S4.3}

We examine ten 52.5 pc radius apertures within the observed region of
IC\,5152.  The extracted PAH/PAH radiance ratios are presented in
Table \ref{ratios5152_1}. Of the 50 ratios shown in that table, only 1
(the (7.7 $\mu$m)/(6.2 $\mu$m) ratio for Region\,5) shows a deviance from
the weighted mean at even the 2$\sigma$ level.  Similarly, as shown in
Table~\ref{ratios5152_2}, there are few variations of the ratios of
[Ne~II] to PAH emission, or of the ratios of 24 $\mu$m to PAH
emission, that are statistically significant.

The constancy of these ratios again arises from a wide variety of
local conditions within IC\,5152.  In the most extreme example in our
limited sample, the IR, optical, and UV morphologies are decidedly
different in IC\,5152.  Most of the UV clusters have associated IR
emission, but the converse in not true; a significant IR component
exists in this system that has no associated UV, optical or
near-infrared emission.  Similarly, the dominant cluster in the UV,
optical and near-infrared has very little associated PAH or dust
emission.  This dichotomy has been seen previously in nearby dwarfs
\citep[e.g.,][]{Cannon06a} and is suggestive of a substantial embedded
star-forming population in IC\,5152.

Considering the regions with extracted radiances, Region\,10 contains
the brightest 24 $\mu$m source in the galaxy; it is coincident with a
UV cluster and is bright in the 8 $\mu$m and 11.3 $\mu$m PAH bands, as
well as luminous in the [Ne~II] emission line.  While its 24 $\mu$m to
PAH ratios are high, the ratios of the PAH emission features
themselves are indistinguishable from other regions in IC\,5152.
Region\,8 has high surface brightnesses in each of the panels of
Figure~\ref{contour5152}, including in the [Ne~II] emission line.
Again, its PAH/PAH ratios appear to be average compared to others
within the galaxy.  At the opposite extreme is Region\,6, which
samples a region of low IR surface brightness.  While diffuse UV
emission is coincident with this aperture, its PAH/PAH radiance ratios
again appear normal.

Taken as a whole, the apertures in IC\,5152 lead to the same
conclusions as those for NGC\,55 and NGC\,3109.  While substantial
local environmental changes occur within the dwarf galaxies of this
limited sample, those changes do not produce measurable differences in
the intensities of the major PAH bands.  The radiance ratios of these
prominent MIR features appear to be insensitive to a wide variety of
local environmental factors.

\section{Conclusions}
\label{S5}

Three nearby (D $\lsim$ 2 Mpc), low-metallicity (12\% Z$_{\odot}$
$\le$ Z $\le$ 49\% Z$_{\odot}$) dwarf galaxies (NGC\,55, NGC\,3109,
and IC\,5152) were observed using the {\it Spitzer}/IRS Short-Low
module in spectral mapping mode.  We examine the resulting spectra on
both global (integrated over the entire field of view, which targets
the regions of high IR surface brightness) and spatially resolved
scales.  These data allow us to study the relative intensities of the
four prominent PAH bands at 6.2 $\mu$m, 7.7 $\mu$m, 8.6 $\mu$m and
11.3 $\mu$m.

Investigation of the global properties of our galaxies reveals the
complex interactions that influence the intensity of PAH emission in
galaxies. While metallicity does correlate with PAH emission (with a
transition metallicity occurring at $\sim$25\% Z$_{\odot}$), our
examination of these three galaxies provides evidence that other
global parameters (e.g., mass, metallicity, current SFR) also affect
the nature of PAH emission.

Over the full area of our spectral maps, the relative ratios of most
PAH bands vary by a factor of 3-4 across our sample.  The exception is
the ratio of (8.6 $\mu$m)/(11.3 $\mu$m) radiances, which is constant on the
global level for all three systems.  Given that the radiation field
strength shows considerable variation across this sample (with the
comparatively metal-rich ISM of NGC\,55 showing widespread [Ne~II]
emission, and the metal-poor ISM of NGC\,3109 having only a few
isolated regions of ionized gas), we interpret the constant value of
the (8.6 $\mu$m)/(11.3 $\mu$m) PAH ratio as evidence for a similar PAH size
distribution across our sample. It then follows that if the PAH size
distribution is similar across our sample, that the grain size
distribution must be fairly long-lived and stable.

We compare the global PAH radiance ratios in our sample galaxies with
a collection of similar measurements that probe a wide variety of
sources from \citet{Smith07a}, \citet{Galliano08}, \citet{Gordon08},
and \citet{Hunt10}. From this comparison, we see evidence for a
similar PAH size distribution across the range of objects explored,
which includes spirals, AGN, dwarfs, and H~II regions. The ionization
indices vary significantly within these different sources. Taken
together, these properties suggest that the local environment has a
greater impact on the intensity of PAH emission than does the
metallicity or the history of PAH formation/destruction in a given
galaxy or region.

We also examine PAH emission on a spatially resolved basis by
extracting maps of PAH features and emission lines in our sample
galaxies.  Examining both individual pixels (in regions of high S/N)
and apertures of physical radius 52.5 pc, we find that the
(7.7 $\mu$m)/(11.3 $\mu$m) ratio (a probe of ionization index or radiation
field strength) varies by a factor of $\sim$5 across our sample.  The
grain size distribution (using the (8.6 $\mu$m)/(6.2 $\mu$m) ratio as a
diagnostic) shows only very minor variations over this range.  This
agrees with the interpretations found for other galaxies and suggests
that there is very little change in the PAH carrier size distribution
as a function of radiation field strength.

Previous works have suggested that the relative intensities of the
main PAH features are essentially constant within a given galaxy (see
references in \S~\ref{S3} and \ref{S4} above).  Our examination of the
ISM in the three dwarf galaxies of this sample supports this
interpretation.  In comparing the PAH/PAH ratios at a spatial
resolution of 52.5 pc, we find very few regions that display
statistically significant deviations from the mean within a given
system (no variations exceed the 3.5$\sigma$ level, and most are at
the 2$\sigma$ level or lower). In contrast to the diverse
environmental variations seen between regions in our sample, the PAH
band ratios are constant within each galaxy.

The apparently simple conclusions of this work mask a great deal of
complexity in the canonically ``simple'' ISM of dwarf galaxies.  The
uniformity of metal abundance in these systems is well-documented but
has not yet been explained.  The present work suggests that this
uniformity also extends to the properties of the carriers of the PAH
bands.  When examining each of these systems, we find a remarkably
wide variety of physical conditions: some star formation regions are
UV-bright while others are deeply embedded; some regions have
widespread ionizing photons while others are apparently quiescent;
some regions are dominated by the red stellar continuum while others
have SEDs that rise steeply toward the far-IR.  In response to these
diverse local conditions, the intensity of the PAH emission features
do in fact change markedly, in line with expectations based on the
strength of the local radiation field and the metallicity.  However,
the relative intensities of these PAH bands appear to be strikingly
uniform over these variable local conditions.

\acknowledgements

This work is based on observations made with the {\it Spitzer Space
Telescope}, which is operated by the Jet Propulsion Laboratory,
California Institute of Technology, under a contract with
NASA. Support for this work was provided by NASA through contract
1321212, issued by JPL/Caltech to J.M.C. at Macalester College. RDG
was supported in part by NASA through contracts 1256406 and 1215746
issued by JPL/Caltech to the University of Minnesota. This research
has made use of the NASA/IPAC Extragalactic Database (NED) which is
operated by the Jet Propulsion Laboratory, California Institute of
Technology, under contract with the National Aeronautics and Space
Administration, and NASA's Astrophysics Data System. This publication
has made use of data products from the Two Micron All Sky Survey,
which is a joint project of the University of Massachusetts and the
Infrared Processing and Analysis Center/California Institute of
Technology, funded by the National Aeronautics and Space
Administration and the National Science Foundation.  We would like to
acknowledge Daniel A. Dale, J.D. Smith, Thomas Varberg, and the
{\it Spitzer} Science Center for helpful discussions and support.  Finally,
we thank the anonymous referee for a careful and insightful report
that improved this manuscript.

\clearpage
\bibliographystyle{apj}

\begin{thebibliography}{}

\bibitem[Allamandola \etal(1999)]{Allamandola99} Allamandola, L.~J.,
Hudgins, D.~M., \& Sandford, S.~A.\ 1999, \apjl, 511, L115

\bibitem[Allamandola \etal(1989)]{Allamandola89} Allamandola, L.~J.,
Tielens, A.~G.~G.~M., \& Barker, J.~R.\ 1989, \apjs, 71, 733

\bibitem[Bakes \etal(2001)]{Bakes01} Bakes, E.~L.~O., Tielens,
A.~G.~G.~M., \& Bauschlicher, C.~W., Jr.\ 2001, \apj, 556, 501

\bibitem[Bauschlicher(2002)]{Bauschlicher02} Bauschlicher, C.~W., Jr.\ 
2002, \apj, 564, 782 

\bibitem[Bauschlicher \etal(2009)]{Bauschlicher09} Bauschlicher, 
C.~W., Peeters, E., \& Allamandola, L.~J.\ 2009, \apj, 697, 311 

\bibitem[Bauschlicher \etal(2008)]{Bauschlicher08} Bauschlicher, 
C.~W., Jr., Peeters, E., \& Allamandola, L.~J.\ 2008, \apj, 678, 316

\bibitem[Cannon \etal(2006)]{Cannon06a} Cannon, J.~M., \etal\ 
2006, \apj, 647, 293 

\bibitem[Cannon \etal(2006)]{Cannon06b} Cannon, J.~M., \etal\ 
2006, \apj, 652, 1170 

\bibitem[Compi{\`e}gne \etal(2007)]{Compiegne07} Compi{\`e}gne, M.,
Abergel, A., Verstraete, L., Reach, W.~T., Habart, E., Smith, J.~D.,
Boulanger, F., \& Joblin, C.\ 2007, \aap, 471, 205

\bibitem[Cox(2000)]{Cox00} Cox, A.~N.\ 2000, Allen's Astrophysical
Quantities

\bibitem[Dale \etal(2009)]{Dale09a} Dale, D.~A., \etal\ 2009, 
\apj, 693, 1821 

\bibitem[Draine \& Li(2007)]{Draine07} Draine, B.~T., \& Li, A.\ 2007,
\apj, 657, 810

\bibitem[Engelbracht \etal(2005)]{Engelbracht05} Engelbracht, C.~W.,
Gordon, K.~D., Rieke, G.~H., Werner, M.~W., Dale, D.~A., \& Latter,
W.~B.\ 2005, \apjl, 628, L29

\bibitem[Engelbracht \etal(2008)]{Engelbracht08} Engelbracht, C.~W.,
Rieke, G.~H., Gordon, K.~D., Smith, J.-D.~T., Werner, M.~W.,
Moustakas, J., Willmer, C.~N.~A., \& Vanzi, L.\ 2008, \apj, 678, 804

\bibitem[Galliano \etal(2008)]{Galliano08} Galliano, F., Madden,
S.~C., Tielens, A.~G.~G.~M., Peeters, E., \& Jones, A.~P.\ 2008, \apj,
679, 310

\bibitem[Giard \etal(1994)]{Giard94} Giard, M., Lamarre, J.~M., Pajot,
F., \& Serra, G.\ 1994, \aap, 286, 203

\bibitem[Gordon \etal(2008)]{Gordon08} Gordon, K.~D., Engelbracht,
C.~W., Rieke, G.~H., Misselt, K.~A., Smith, J.-D.~T., \& Kennicutt,
R.~C., Jr.\ 2008, \apj, 682, 336

\bibitem[Helou \etal(2000)]{Helou00} Helou, G., Lu, N.~Y., 
Werner, M.~W., Malhotra, S., \& Silbermann, N.\ 2000, \apjl, 532, L21 


\bibitem[Helou \etal(2001)]{Helou01} Helou, G., Malhotra, S.,
Hollenbach, D.~J., Dale, D.~A., \& Contursi, A.\ 2001, \apjl, 548, L73

\bibitem[Helou \etal(2004)]{Helou04} Helou, G., \etal\ 2004, \apjs,
154, 253

\bibitem[Hogg \etal(2005)]{Hogg05} Hogg, D.~W., Tremonti, 
C.~A., Blanton, M.~R., Finkbeiner, D.~P., Padmanabhan, N., Quintero, A.~D., 
Schlegel, D.~J., \& Wherry, N.\ 2005, \apj, 624, 162 

\bibitem[Houck \etal(2004)]{Houck04} Houck, J.~R., \etal\ 2004,
\apjs, 154, 18

\bibitem[Hunt \etal(2010)]{Hunt10} Hunt, L.~K., Thuan, T.~X., Izotov,
Y.~I., \& Sauvage, M.\ 2010, \apj, 712, 164

\bibitem[Jackson \etal(2006)]{Jackson06} Jackson, D.~C., Cannon,
J.~M., Skillman, E.~D., Lee, H., Gehrz, R.~D., Woodward, C.~E., \&
Polomski, E.\ 2006, \apj, 646, 192

\bibitem[Jarrett \etal(2003)]{jarrett03} Jarrett, T.~H., Chester, T.,
  Cutri, R., Schneider, S.~E., \& Huchra, J.~P.\ 2003, \aj, 125, 525

\bibitem[Kennicutt \etal(2003)]{Kennicutt03} Kennicutt, R.~C., Jr.,
\etal\ 2003, \pasp, 115, 928

\bibitem[Kim \& Saykally(2002)]{Kim02} Kim, H.-S., \& Saykally, R.~J.\
2002, \apjs, 143, 455

\bibitem[Kobulnicky \& Skillman(1996)]{Kobulnicky96} Kobulnicky,
H.~A., \& Skillman, E.~D.\ 1996, \apj, 471, 211

\bibitem[Kobulnicky \& Skillman(1997)]{Kobulnicky97} Kobulnicky,
H.~A., \& Skillman, E.~D.\ 1997, \apj, 489, 636

\bibitem[Lee \etal(2003a)]{Lee03a} Lee, H., Grebel, E.~K., \& Hodge,
P.~W.\ 2003, \aap, 401, 141

\bibitem[Lee \etal(2003b)]{Lee03b} Lee, H., McCall, M.~L., Kingsburgh,
R.~L., Ross, R., \& Stevenson, C.~C.\ 2003, \aj, 125, 146

\bibitem[Lee \etal(2009)]{Lee09} Lee, J.~C., \etal\ 2009, \apj, 706,
599

\bibitem[Leger \& Puget(1984)]{Leger84} Leger, A., \& Puget, J.~L.\
1984, \aap, 137, L5

\bibitem[Leroy \etal(2005)]{Leroy05} Leroy, A., Bolatto, A.~D.,
Simon, J.~D., \& Blitz, L.\ 2005, \apj, 625, 763

\bibitem[Mateo(1998)]{Mateo98} Mateo, M.~L.\ 1998, \araa, 36, 435

\bibitem[O'Halloran \etal(2006)]{Ohalloran06} O'Halloran, B., 
Satyapal, S., \& Dudik, R.~P.\ 2006, \apj, 641, 795 

\bibitem[Ricks \etal(2009)]{Ricks09} Ricks, A.~M., Douberly, 
G.~E., \& Duncan, M.~A.\ 2009, \apj, 702, 301 

\bibitem[Rosenberg \etal(2006)]{Rosenberg06} Rosenberg, J.~L., Ashby,
M.~L.~N., Salzer, J.~J., \& Huang, J.-S.\ 2006, \apj, 636, 742

\bibitem[Sanders \& Mirabel(1996)]{Sanders96} Sanders, D.~B., \&
Mirabel, I.~F.\ 1996, \araa, 34, 749

\bibitem[Sanders \etal(1988)]{Sanders88} Sanders, D.~B., Soifer,
B.~T., Elias, J.~H., Madore, B.~F., Matthews, K., Neugebauer, G., \&
Scoville, N.~Z.\ 1988, \apj, 325, 74

\bibitem[Skrutskie \etal(2006)]{skrutskie} Skrutskie, M.~F., et 
al.\ 2006, \aj, 131, 1163

\bibitem[Smith \etal(2007)]{Smith07b} Smith, J.~D.~T., \etal\ 
2007b, \pasp, 119, 1133 

\bibitem[Smith \etal(2007)]{Smith07a} Smith, J.~D.~T., \etal\ 
2007a, \apj, 656, 770 

\bibitem[Taylor \etal(1998)]{Taylor98} Taylor, C.~L., Kobulnicky,
H.~A., \& Skillman, E.~D.\ 1998, \aj, 116, 2746

\bibitem[Tenorio-Tagle \& Bodenheimer(1988)]{Tenoriotagle88}
Tenorio-Tagle, G., \& Bodenheimer, P.\ 1988, \araa, 26, 145

\bibitem[Tielens(2008)]{Tielens08} Tielens, A.~G.~G.~M.\ 2008, \araa,
46, 289

\bibitem[Webster \& Smith(1983)]{Webster83} Webster, B.~L., \& Smith,
M.~G.\ 1983, \mnras, 204, 743

\bibitem[Wu \etal(2010)]{Wu10} Wu, R., Hogg, D.~W., \& Moustakas, J.
2010, \apj, submiited (arxiv/0907.1783)

\end{thebibliography}


\clearpage
\begin{deluxetable}{lllccccc}
\rotate 
\tablecaption{Galaxy Sample Properties}
\tablewidth{0pt}
\tablehead{\colhead{Target}                           & 
           \colhead{Type \tablenotemark{a}}           &
           \colhead{M$_B$}                            & 
           \colhead{Distance\tablenotemark{b}}        &
           \colhead{12+log(O/H)\tablenotemark{c}}     & 
           \colhead{Mass}                             &
           \colhead{Current Star}                     & 
           \colhead{Diffuse 8 $\mu$m}                        \\ 
           \colhead{}                                 &
           \colhead{}                                 & 
           \colhead{}                                 & 
           \colhead{}                                 & 
           \colhead{}                                 &  
           \colhead{}                                 &
           \colhead{Formation Rate}                   &
           \colhead{Flux Density\tablenotemark{d}} \\ 
           \colhead{}                                 & 
           \colhead{}                                 & 
           \colhead{(mag)}                            & 
           \colhead{(Mpc)}                            &
           \colhead{(dex)}                            & 
           \colhead{(10$^6$ M$_{\odot}$)}              &
           \colhead{(M$_{\odot}$ yr$^{-1}$)}            & 
           \colhead{(Jy)}                              }                          
\startdata 
NGC\,55 & IrrIV & -17.5 & 2.17 & 8.4\,$\pm$\,0.1 & 15600 & 5.40$\times$10$^{-1}$ & 1.4\\ 
NGC\,3109 & IrrIV-V & -15.2 & 1.34 & 7.74\,$\pm$\,0.33 & 6550 & 3.98$\times$10$^{-2}$ & 0.06 \\ 
IC\,5152 & dIrr & -14.5 & 1.97 & 7.92\,$\pm$\,0.07 & 400 & 3.47$\times$10$^{-2}$ & 0.16 \\ 
\enddata 
\tablenotetext{a}{Type,
  M$_B$, and mass (calculated from central or rotational velocity, as available) taken from \citet{Mateo98}}
\tablenotetext{b}{Distance and SFR taken from \citet{Lee09}. SFR based
  on GALEX FUV integrated photometry and corrected for extinction
  based on A$_{FUV}$ = 7.9E(B -- V), except for NGC\,3109, where UV
  attenuation was unavailable, and H$\alpha$ attenuation
  (A$_{H\alpha}$ = 2.5E(B -- V), scaled to a factor of 1.8, was used
  instead} 
\tablenotetext{c}{Metallicities taken from \citet{Webster83} for NGC\,55, \citet{Lee03a} for IC\,5152, and \citet{Lee03b} for NGC\,3109.}
\tablenotetext{d}{Diffuse 8 $\mu$m flux density taken from \citet{Jackson06}}
\label{galaxysample}
\end{deluxetable}

\clearpage
\begin{deluxetable}{lccc}
\tablecaption{Global Feature Radiances From PAHFIT\tablenotemark{a}}
\tablewidth{0pt}
\tablehead{\colhead{Wavelength}   &
           \colhead{NGC\,55}       &
           \colhead{NGC\,3109}     &
           \colhead{IC\,5152}      }
\startdata
6.2 $\mu$m PAH & $ 76.5 \pm 0.9  $   & $ 4.5  \pm 0.6 $   & $ 12.0  \pm 0.7 $   \\
7.7 $\mu$m PAH & $ 226.1  \pm 3.6  $   & $ 12.8  \pm 1.9 $   & $ 26.3  \pm 2.7 $   \\
8.6 $\mu$m PAH & $ 43.0 \pm 0.6  $   & $ 0.9  \pm 0.4 $   & $ 8.4  \pm 0.3 $   \\
$ [$S IV$]$ 10.5 $\mu$m        & $ 7.0 \pm 0.1 $   & $ 0.2  \pm 0.1 $  & $ 0.5  \pm 0.1 $  \\
11.3 $\mu$m PAH & $ 73.4 \pm 0.9  $   & $ 1.9  \pm 0.2 $   & $ 14.3  \pm 0.5 $   \\
12.6 $\mu$m PAH & $ 29.7 \pm 0.4  $   & $ 0.9  \pm 0.1 $   & $ 3.2  \pm 0.3 $   \\
$[$Ne~II$]$ 12.8 $\mu$m       & $ 8.3 \pm 0.1 $   & $ 0.5  \pm 0.1 $  & $ 1.4  \pm 0.1 $  \\
\enddata
\label{globalpahline}
\tablenotetext{a}{Units of 10$^{-9}$ W m$^{-2}$ sr$^{-1}$.}
\end{deluxetable}

\clearpage
\begin{deluxetable}{lccccc}
\rotate
\tablecaption{Global PAH Ratios\tablenotemark{a}}
\tablewidth{0pt}
\tablehead{\colhead{Target}                               &
           \colhead{(6.2 $\mu$m)/(11.3 $\mu$m)}                 &
           \colhead{(7.7 $\mu$m)/(11.3 $\mu$m)}                 &
           \colhead{(8.6 $\mu$m)/(11.3 $\mu$m)}                 &
           \colhead{(7.7 $\mu$m)/(6.2 $\mu$m)}                  &
           \colhead{(8.6 $\mu$m)/(6.2 $\mu$m)}                  }
\startdata
NGC\,55   & 1.04 $\pm$ 0.02 & 3.07 $\pm$ 0.06 & 0.59 $\pm$ 0.01 & 2.95 $\pm$ 0.06 & 0.56 $\pm$ 0.01  \\
NGC\,3109 & 2.39 $\pm$ 0.38 & 6.72 $\pm$ 1.14 & 0.46 $\pm$ 0.19 & 2.81 $\pm$ 0.56 & 0.19 $\pm$ 0.08  \\
IC\,5152  & 0.84 $\pm$ 0.06 & 1.84 $\pm$ 0.20 & 0.58 $\pm$ 0.03 & 2.20 $\pm$ 0.25 & 0.70 $\pm$ 0.05  \\
\enddata
\label{globalcomp}
\tablenotetext{a}{Values are ratios of radiances measured in W m$^{-2}$ sr$^{-1}$.}
\end{deluxetable}

\clearpage
\begin{deluxetable}{lcccccc}
\rotate
\tablecaption{Spectral Extraction Wavelength Regions\tablenotemark{a}}
\tablewidth{0pt}
\tablehead{\colhead{Target}               &
           \colhead{6.2 $\mu$m}            &
           \colhead{7.7 $\mu$m}            &
           \colhead{8.6 $\mu$m}             &
           \colhead{Broad 8 $\mu$m}        &
           \colhead{11.3 $\mu$m}           &          
           \colhead{[Ne~II] 12.8 $\mu$m}   }
\startdata
NGC\,55   & 5.986-6.513 & 7.536-7.882 & 8.403-8.838 & 7.288-8.900 & 10.949-11.694 & 12.625-12.998 \\
NGC\,3109 & 6.141-6.358 & 7.443-7.782 & 8.217-8.900 & 7.288-8.900 & 10.763-11.570 & 12.688-12.936 \\
IC\,5152  & 6.110-6.420 & 7.474-7.882 & 8.403-8.838 & 7.288-8.900 & 11.073-11.446 & 12.688-12.998 \\
\enddata
\label{extract}
\tablenotetext{a}{Spectral extraction regions are given in units of $\mu$m.}
\end{deluxetable}

\clearpage
\begin{deluxetable}{cccccc}
\rotate
\tablecaption{PAH/PAH Ratios for NGC\,55 Extraction Regions\tablenotemark{a}}
\tablewidth{0pt}
\tablehead{\colhead{Region} &
           \colhead{(6.2 $\mu$m)/(11.3 $\mu$m)} &
           \colhead{(7.7 $\mu$m)/(11.3 $\mu$m)} &
           \colhead{(8.6 $\mu$m)/(11.3 $\mu$m)} &
           \colhead{(7.7 $\mu$m)/(6.2 $\mu$m)} &
           \colhead{(8.6 $\mu$m)/(6.2 $\mu$m)} }
\startdata
Region 1  & 0.93 $\pm$ 0.07 & 2.73 $\pm$ 0.29 & 0.53 $\pm$ 0.04 & 2.92 $\pm$ 0.36 & 0.56 $\pm$ 0.06 \\
Region 2  & 1.15 $\pm$ 0.16 & 2.41 $\pm$ 0.50 & 0.34 $\pm$ 0.08 & 2.10 $\pm$ 0.51 & 0.29 $\pm$ 0.08 \\
Region 3  & 1.01 $\pm$ 0.07 & 2.97 $\pm$ 0.21 & 0.45 $\pm$ 0.04 & 2.93 $\pm$ 0.27 & 0.44 $\pm$ 0.05 \\
Region 4  & 1.14 $\pm$ 0.08 & 3.28 $\pm$ 0.27 & 0.55 $\pm$ 0.04 & 2.87 $\pm$ 0.29 & 0.48 $\pm$ 0.04 \\
Region 5  & 1.04 $\pm$ 0.05 & 2.88 $\pm$ 0.15 & 0.56 $\pm$ 0.03 & 2.78 $\pm$ 0.14 & 0.54 $\pm$ 0.02 \\
Region 6  & 1.10 $\pm$ 0.05 & 3.29 $\pm$ 0.14 & 0.59 $\pm$ 0.03 & 2.99 $\pm$ 0.12 & 0.54 $\pm$ 0.03 \\
Region 7  & 1.02 $\pm$ 0.04 & 3.07 $\pm$ 0.11 & 0.51 $\pm$ 0.02 & 3.01 $\pm$ 0.16 & 0.50 $\pm$ 0.03 \\
Region 8  & 1.10 $\pm$ 0.05 & 3.28 $\pm$ 0.16 & 0.57 $\pm$ 0.03 & 2.98 $\pm$ 0.12 & 0.52 $\pm$ 0.02 \\
Region 9  & 1.32 $\pm$ 0.04 & 3.88 $\pm$ 0.11 & 0.63 $\pm$ 0.02 & 2.94 $\pm$ 0.09 & 0.47 $\pm$ 0.02 \\
Region 10 & 1.12 $\pm$ 0.06 & 2.88 $\pm$ 0.17 & 0.53 $\pm$ 0.03 & 2.57 $\pm$ 0.20 & 0.48 $\pm$ 0.04 \\
Region 11 & 1.11 $\pm$ 0.05 & 3.58 $\pm$ 0.15 & 0.60 $\pm$ 0.03 & 3.23 $\pm$ 0.12 & 0.55 $\pm$ 0.03 \\
Region 12 & 1.15 $\pm$ 0.04 & 3.33 $\pm$ 0.12 & 0.57 $\pm$ 0.02 & 2.91 $\pm$ 0.13 & 0.50 $\pm$ 0.02 \\
Region 13 & 1.14 $\pm$ 0.03 & 3.51 $\pm$ 0.12 & 0.57 $\pm$ 0.02 & 3.09 $\pm$ 0.13 & 0.50 $\pm$ 0.02 \\
Region 14 & 1.12 $\pm$ 0.05 & 3.56 $\pm$ 0.17 & 0.62 $\pm$ 0.03 & 3.17 $\pm$ 0.15 & 0.55 $\pm$ 0.03 \\
Region 15 & 1.11 $\pm$ 0.08 & 3.20 $\pm$ 0.22 & 0.49 $\pm$ 0.04 & 2.88 $\pm$ 0.19 & 0.45 $\pm$ 0.04 \\
Region 16 & 1.06 $\pm$ 0.14 & 2.31 $\pm$ 0.19 & 0.37 $\pm$ 0.07 & 2.17 $\pm$ 0.32 & 0.35 $\pm$ 0.08 \\
Region 17 & 1.13 $\pm$ 0.05 & 3.20 $\pm$ 0.20 & 0.59 $\pm$ 0.04 & 2.82 $\pm$ 0.18 & 0.52 $\pm$ 0.03 \\
Region 18 & 1.23 $\pm$ 0.12 & 2.66 $\pm$ 0.39 & 0.47 $\pm$ 0.07 & 2.16 $\pm$ 0.37 & 0.38 $\pm$ 0.06 \\
Region 19 & 1.27 $\pm$ 0.11 & 2.66 $\pm$ 0.39 & 0.47 $\pm$ 0.06 & 2.10 $\pm$ 0.32 & 0.37 $\pm$ 0.05 \\
\hline
Weighted Mean      & 1.13 $\pm$ 0.01 & 3.27 $\pm$ 0.04 & 0.56 $\pm$ 0.01 & 2.94 $\pm$ 0.04 & 0.50 $\pm$ 0.01 \\
Standard Deviation & 0.09 & 0.42 & 0.08 & 0.37 & 0.08 \\
\enddata
\label{ratios55_1}
\tablenotetext{a}{Values are ratios of radiances measured in of W
  m$^{-2}$ sr$^{-1}$.}
\end{deluxetable}

\clearpage
\begin{deluxetable}{ccccc}
\rotate
\tablecaption{Selected Ratios for NGC\,55 Extraction Regions\tablenotemark{a}}
\tablewidth{0pt}
\tablehead{\colhead{Region} &
           \colhead{$[$Ne~II$]$/(11.3 $\mu$m)} &
           \colhead{$[$Ne~II$]$/(8.6 $\mu$m)} &
           \colhead{(MIPS\,24 $\mu$m)/(11.3 $\mu$m)} &
           \colhead{(MIPS\,24 $\mu$m)/(8.6 $\mu$m)} }
\startdata
Region 1  & 0.03 $\pm$ 0.01 \tablenotemark{b} & 0.05 $\pm$ 0.01 & 59.7 $\pm$ 1.48 & 113 $\pm$ 9.13 \\
Region 2  & 0.06 $\pm$ 0.01 & 0.16 $\pm$ 0.05 & 191 $\pm$ 8.08 & 563 $\pm$ 131 \\
Region 3  & 0.10 $\pm$ 0.01 & 0.23 $\pm$ 0.02 & 246 $\pm$ 4.80 & 548 $\pm$ 44.4 \\
Region 4  & 0.10 $\pm$ 0.01 & 0.18 $\pm$ 0.02 & 110 $\pm$ 2.26 & 198 $\pm$ 13.1 \\
Region 5  & 0.11 $\pm$ 0.01 & 0.19 $\pm$ 0.01 & 97.6 $\pm$ 3.25 &  174 $\pm$ 6.09 \\
Region 6  & 0.10 $\pm$ 0.01 & 0.17 $\pm$ 0.01 & 101 $\pm$ 3.45 & 170 $\pm$ 6.08 \\
Region 7 & 0.05 $\pm$ 0.01 & 0.10 $\pm$ 0.01 & 66.2 $\pm$ 0.67 & 129 $\pm$ 4.73 \\
Region 8 & 0.07 $\pm$ 0.01 & 0.12 $\pm$ 0.01 & 95.1 $\pm$ 3.56 & 166 $\pm$ 5.98 \\
Region 9  & 0.15 $\pm$ 0.01 & 0.23 $\pm$ 0.01 & 333 $\pm$ 5.25 & 530 $\pm$ 14.2 \\
Region 10 & 0.04 $\pm$ 0.01 & 0.08 $\pm$ 0.01 & 905 $\pm$ 15.01 & 1700 $\pm$ 105 \\
Region 11 & 0.08 $\pm$ 0.01 & 0.13 $\pm$ 0.01 & 101 $\pm$ 3.45 & 168 $\pm$ 6.38 \\
Region 12  & 0.06 $\pm$ 0.01 & 0.11 $\pm$ 0.01 & 114 $\pm$ 1.38 & 199 $\pm$ 7.85 \\
Region 13  & 0.09 $\pm$ 0.01 & 0.15 $\pm$ 0.01 & 106 $\pm$ 1.07 & 185 $\pm$ 5.82 \\
Region 14 & 0.10 $\pm$ 0.01 & 0.16 $\pm$ 0.01 & 178 $\pm$ 5.75 & 287 $\pm$ 11.5 \\
Region 15 & 0.09 $\pm$ 0.01 & 0.17 $\pm$ 0.01 & 157 $\pm$ 8.09 & 317 $\pm$ 20.8 \\
Region 16 & 0.03 $\pm$ 0.01 & 0.08 $\pm$ 0.03 & 689 $\pm$ 25.51 & 1870 $\pm$ 369 \\
Region 17 & 0.17 $\pm$ 0.01 & 0.29 $\pm$ 0.02 & 221 $\pm$ 7.01 & 377 $\pm$ 19.6 \\
Region 18 & 0.05 $\pm$ 0.01 & 0.11 $\pm$ 0.03 & 521 $\pm$ 16.51 & 1100 $\pm$ 152 \\
Region 19 & 0.17 $\pm$ 0.01 & 0.36 $\pm$ 0.05 & 701 $\pm$ 34.90 & 1490 $\pm$ 175 \\
\hline
Weighted Mean      & 0.08 $\pm$ 0.01 & 0.15 $\pm$ 0.01 & 89.5 $\pm$ 0.46 & 181 $\pm$ 2.07 \\ 
Standard Deviation & 0.04 & 0.08 & 252 & 564 \\
\enddata
\label{ratios55_2}
\tablenotetext{a}{Values are ratios of radiances measured in W m$^{-2}$ sr$^{-1}$.}
\tablenotetext{b}{For some regions, errors were below .005. In these cases, the error was rounded up to 0.01.}
\end{deluxetable}

\clearpage
\begin{deluxetable}{cccccc}
\rotate
\tablecaption{PAH/PAH Ratios for NGC\,3109 Extraction Regions\tablenotemark{a}}
\tablewidth{0pt}
\tablehead{\colhead{Region} &
           \colhead{(6.2 $\mu$m)/(11.3 $\mu$m)} &
           \colhead{(7.7 $\mu$m)/(11.3 $\mu$m)} &
           \colhead{(8.6 $\mu$m)/(11.3 $\mu$m)} &
           \colhead{(7.7 $\mu$m)/(6.2 $\mu$m)} &
           \colhead{(8.6 $\mu$m)/(6.2 $\mu$m)} }
\startdata
Region 1  & 2.45 $\pm$ 1.30 & 6.91 $\pm$ 4.50 & 0.49 $\pm$ 0.66 & 2.82 $\pm$ 1.93 & 0.20 $\pm$ 0.27 \\
Region 2  & 3.09 $\pm$ 1.60 & 9.32 $\pm$ 4.97 & 0.76 $\pm$ 0.79 & 3.02 $\pm$ 1.50 & 0.25 $\pm$ 0.25 \\
Region 3  & 1.55 $\pm$ 0.41 & 5.26 $\pm$ 1.11 & 0.32 $\pm$ 0.24 & 3.40 $\pm$ 0.98 & 0.21 $\pm$ 0.16 \\
Region 4  & 4.01 $\pm$ 1.42 & 10.23 $\pm$ 3.45 & 0.98 $\pm$ 0.48 & 2.55 $\pm$ 0.89 & 0.24 $\pm$ 0.12 \\
\hline
Weighted Mean      & 1.86 $\pm$ 0.37 & 5.93 $\pm$ 1.00 & 0.47 $\pm$ 0.19 & 2.94 $\pm$ 0.58 & 0.23 $\pm$ 0.09 \\
Standard Deviation & 1.04 & 2.26 & 0.29 & 0.36 & 0.02 \\
\enddata
\label{ratios3109_1}
\tablenotetext{a}{Values are ratios of radiances measured in W m$^{-2}$ sr$^{-1}$.}
\end{deluxetable}

\clearpage
\begin{deluxetable}{ccccc}
\rotate
\tablecaption{Selected Ratios for NGC\,3109 Extraction Regions\tablenotemark{a}}
\tablewidth{0pt}
\tablehead{\colhead{Region} &
           \colhead{$[$Ne~II$]$/(11.3 $\mu$m)} &
           \colhead{$[$Ne~II$]$/(8.6 $\mu$m)} &
           \colhead{(MIPS\,24 $\mu$m)/(11.3 $\mu$m)} &
           \colhead{(MIPS\,24 $\mu$m)/(8.6 $\mu$m)} }
\startdata
Region 1  & 0.02 $\pm$ 0.02 & 0.06 $\pm$ 0.10 & 2420 $\pm$ 1420 & 7560 $\pm$ 2550 \\
Region 2  & 0.14 $\pm$ 0.10 & 0.18 $\pm$ 0.20 & 635 $\pm$ 609 & 808 $\pm$ 1850 \\
Region 3  & 0.25 $\pm$ 0.05 & 1.25 $\pm$ 0.03 & 269 $\pm$ 42.4 & 1330 $\pm$ 1950 \\
Region 4  & 0.11 $\pm$ 0.06 & 0.09 $\pm$ 0.06 & 617 $\pm$ 257 & 533 $\pm$ 330 \\
\hline
Weighted Mean      & 0.06 $\pm$ 0.02 & 0.09 $\pm$ 0.05 & 281 $\pm$ 41.7 & 564 $\pm$ 320 \\
Standard Deviation & 0.10 & 0.57 & 972 & 3350 \\
\enddata
\label{ratios3109_2}
\tablenotetext{a}{Values are ratios of radiances measured in of W m$^{-2}$ sr$^{-1}$.}
\end{deluxetable}

\clearpage
\begin{deluxetable}{cccccc}
\rotate
\tablecaption{PAH/PAH Ratios for IC\,5152 Extraction Regions\tablenotemark{a}}
\tablewidth{0pt}
\tablehead{\colhead{Region} &
           \colhead{(6.2 $\mu$m)/(11.3 $\mu$m)} &
           \colhead{(7.7 $\mu$m)/(11.3 $\mu$m)} &
           \colhead{(8.6 $\mu$m)/(11.3 $\mu$m)} &
           \colhead{(7.7 $\mu$m)/(6.2 $\mu$m)} &
           \colhead{(8.6 $\mu$m)/(6.2 $\mu$m)} }
\startdata
Region 1  & 1.18 $\pm$ 0.47 & 2.63 $\pm$ 1.58 & 0.36 $\pm$ 0.26 & 2.22 $\pm$ 1.50 & 0.30 $\pm$ 0.24 \\
Region 2  & 1.25 $\pm$ 0.14 & 3.17 $\pm$ 0.53 & 0.58 $\pm$ 0.10 & 2.52 $\pm$ 0.48 & 0.46 $\pm$ 0.09 \\
Region 3  & 0.96 $\pm$ 0.10 & 3.04 $\pm$ 0.39 & 0.41 $\pm$ 0.06 & 3.18 $\pm$ 0.50 & 0.43 $\pm$ 0.07 \\
Region 4  & 1.01 $\pm$ 0.17 & 3.08 $\pm$ 0.46 & 0.55 $\pm$ 0.09 & 3.04 $\pm$ 0.65 & 0.54 $\pm$ 0.12 \\
Region 5  & 0.80 $\pm$ 0.20 & 3.89 $\pm$ 0.31 & 0.21 $\pm$ 0.11 & 4.87 $\pm$ 1.23 & 0.26 $\pm$ 0.15 \\
Region 6  & 0.83 $\pm$ 0.33 & 1.80 $\pm$ 0.44 & 0.22 $\pm$ 0.15 & 2.18 $\pm$ 0.84 & 0.27 $\pm$ 0.20 \\
Region 7  & 0.83 $\pm$ 0.11 & 2.41 $\pm$ 0.32 & 0.41 $\pm$ 0.06 & 2.89 $\pm$ 0.54 & 0.49 $\pm$ 0.10 \\
Region 8  & 0.83 $\pm$ 0.17 & 2.39 $\pm$ 0.52 & 0.44 $\pm$ 0.11 & 2.89 $\pm$ 0.58 & 0.53 $\pm$ 0.12 \\
Region 9  & 0.75 $\pm$ 0.16 & 2.15 $\pm$ 0.43 & 0.46 $\pm$ 0.08 & 2.85 $\pm$ 0.81 & 0.61 $\pm$ 0.16 \\
Region 10 & 1.08 $\pm$ 0.17 & 3.17 $\pm$ 0.59 & 0.64 $\pm$ 0.10 & 2.93 $\pm$ 0.53 & 0.59 $\pm$ 0.09 \\
\hline
Weighted Mean      & 0.95 $\pm$ 0.05 & 2.85 $\pm$ 0.14 & 0.45 $\pm$ 0.03 & 2.89 $\pm$ 0.20 & 0.48 $\pm$ 0.04 \\
Standard Deviation & 0.17 & 0.61 & 0.14 & 0.75 & 0.13 \\
\enddata
\label{ratios5152_1}
\tablenotetext{a}{Values are ratios of radiances measured in W m$^{-2}$ sr$^{-1}$.}
\end{deluxetable}

\clearpage
\begin{deluxetable}{ccccc}
\rotate
\tablecaption{Selected Ratios for IC\,5152 Extraction Regions\tablenotemark{a}}
\tablewidth{0pt}
\tablehead{\colhead{Region} &
           \colhead{$[$Ne~II$]$/(11.3 $\mu$m)} &
           \colhead{$[$Ne~II$]$/(8.6 $\mu$m)} &
           \colhead{(MIPS\,24 $\mu$m)/(11.3 $\mu$m)} &
           \colhead{(MIPS\,24 $\mu$m)/(8.6 $\mu$m)}}
\startdata
Region 1  & 0.13 $\pm$ 0.05 & 0.36 $\pm$ 0.28 & 71.6 $\pm$ 17.7 & 201.59 $\pm$ 172 \\
Region 2  & 0.08 $\pm$ 0.01 & 0.14 $\pm$ 0.03 & 48.2 $\pm$ 2.07 & 83.28 $\pm$ 13.3 \\
Region 3  & 0.07 $\pm$ 0.01 & 0.17 $\pm$ 0.03 & 53.2 $\pm$ 1.59 & 128.15 $\pm$ 17.6 \\
Region 4  & 0.07 $\pm$ 0.01 & 0.13 $\pm$ 0.03 & 50.8 $\pm$ 2.10 & 91.87 $\pm$ 13.6 \\
Region 5  & 0.08 $\pm$ 0.03 & 0.38 $\pm$ 0.23 & 43.2 $\pm$ 2.43 & 205.54 $\pm$ 105 \\
Region 6  & 0.06 $\pm$ 0.03 & 0.29 $\pm$ 0.23 & 78.9 $\pm$ 14.4 & 354.78 $\pm$ 236 \\
Region 7  & 0.10 $\pm$ 0.01 & 0.25 $\pm$ 0.05 & 41.7 $\pm$ 1.42 & 102.57 $\pm$ 15.1 \\
Region 8  & 0.06 $\pm$ 0.01 & 0.13 $\pm$ 0.03 & 39.0 $\pm$ 6.04 & 89.01 $\pm$ 17.6 \\
Region 9  & 0.04 $\pm$ 0.02 & 0.10 $\pm$ 0.04 & 41.7 $\pm$ 1.97 & 91.13 $\pm$ 15.7 \\
Region 10 & 0.13 $\pm$ 0.02 & 0.20 $\pm$ 0.03 & 140 $\pm$ 15.8 & 218.37 $\pm$ 24.5 \\
\hline
Weighted Mean      & 0.08 $\pm$ 0.01 & 0.16 $\pm$ 0.01 & 46.6 $\pm$ 0.74 & 104 $\pm$ 6.00 \\
Standard Deviation & 0.03 & 0.10 & 30.7 & 88.0 \\
\enddata
\label{ratios5152_2}
\tablenotetext{a}{Values are ratios of radiances measured in W
  m$^{-2}$ sr$^{-1}$.}
\end{deluxetable}

\clearpage

\clearpage 

\clearpage 

\clearpage 

\clearpage 

\clearpage 

\clearpage 

\clearpage

\clearpage

\clearpage

\clearpage

\clearpage

\clearpage

\clearpage

\clearpage

\clearpage

\clearpage

\end{document}